\documentstyle[aps,psfig]{revtex}

\begin{document}

\draft

\title{Spectral ergodicity and normal modes in ensembles of sparse matrices}

\author{A.D. Jackson,$^1$ C. Mejia-Monasterio,$^2$ T. Rupp,$^3$ 
        M. Saltzer,$^4$ and T. Wilke$^1$}

\address{
	$^1$The Niels Bohr Institute, Blegdamsvej 17, DK-2100
        Copenhagen \O, Danmark\\
        $^2$Centro de Ciencias Fisicas, UNAM, Apartado, 
        Postal 48-3, Cuernavaca, Mexico\\
        $^3$Max-Planck-Institut f\"ur Kernphysik, Postfach 103980, 
        D-69029 Heidelberg, Germany\\ 
        $^4$Albert-Ludwigs-Universit\"at Freiburg,
        Hermann-Herder-Str. 3, D-79104 Freiburg, Germany}

\date{\today}

\maketitle

\begin{abstract}
We investigate the properties of sparse matrix ensembles with
particular regard for the spectral ergodicity hypothesis, which claims
the identity of ensemble and spectral averages of spectral
correlators.  An apparent violation of the spectral ergodicity is
observed.  This effect is studied with the aid of the normal modes of
the random matrix spectrum, which describe fluctuations of the
eigenvalues around their average positions.  This analysis reveals
that spectral ergodicity is not broken, but that different energy
scales of the spectra are examined by the two averaging techniques.
Normal modes are shown to provide a useful complement to traditional
spectral analysis with possible applications to a wide range of
physical systems.
\end{abstract}


\section{Introduction}
\label{sec1}

Random matrix theory (RMT) has shown itself to be useful in modeling a
wide variety of physical systems~\cite{Bro81,Guh98}.  Sparse random
matrices, which are characterized by a considerable fraction of
vanishing or negligible small matrix elements, have attracted
particular interest in recent years.  Applications of sparse matrices
can be found, for instance, in disordered systems, quantum chaotic
systems or many-body systems~\cite{Guh98}. There are few analytical
investigations of sparse matrices, and these are often restricted to
the subclass of random band matrices, i.e.\ matrices where the
off-diagonal elements vanish for a sufficiently large distance from
the main diagonal. More general classes of sparse matrices must be
treated mainly by numerical methods.

In RMT, spectral quantities are usually calculated as averages over
the matrix ensemble.  Experimental data, however, normally consists of
only a single spectrum, and the corresponding quantities must
therefore be calculated as a running average over the energy.
Comparison of experiment and theory is facilitated by the spectral
ergodicity hypothesis for matrix ensembles: An average over an
ensemble of random matrices provides the same result as an average
over energies for a single element of this ensemble in the limit where
the dimension of the matrices is large.  In the case of full
(i.e.~non-sparse) random matrices only, spectral ergodicity can be
proved rigorously~\cite{Guh98,Plu00,Fre78}.  The question of whether a
given matrix ensemble obeys spectral ergodicity is not merely
academic; it has a direct bearing on the ability to compare
experimental results with theoretical expectations.

In the context of disordered mesoscopic systems, the issue of
ergodicity was first raised in Ref.~\cite{Lee87}. Here the ergodicity
hypothesis states that averages over all realizations of the disorder
potential yield the same values for an observable as averages over an
interval variable such as energy or an applied magnetic field. For
such systems -- we think of a single electron in a crystal with random
impurities -- the Hamiltonian is modeled by a sparse random matrix via
a spatial discretization on a lattice. The disorder potential is then
represented by an appropriate choice of the matrix elements as random
variables. The diagonal elements represent the on-site energies; the
off-diagonal elements represent the coupling between the sites.
Assuming that only neighboring sites are coupled, one obtains a
Hamiltonian of sparse matrix form.  One-dimensional systems, e.g.~long
and thin wires, are described by random band
matrices~\cite{Fyo91a,Fyo91b,Fyo92,Fyo93}.  In higher dimensions, the
inevitable presence of side bands renders analytical treatment
difficult.

The question of the ergodicity of mesoscopic disordered systems is
important for the extraction of the Thouless energy from spectral
data. The Thouless energy, $E_{\rm c}$, is an intrinsic energy scale
in disordered systems~\cite{Tho74,Tho75,Tho77} which is essentially
the inverse of the time required for a particle to move diffusively
through the sample as a consequence of multiple scatterings by
impurities. The Thouless energy is measured in terms of the mean level
spacing, $\Delta$, and can be extracted from the fluctuations in the
eigenvalue spectrum of the corresponding Hamiltonian. Here the mean
level spacing $\Delta$ is the inverse of the spectral density, $\rho$,
and is a function of energy.  Furthermore, the Thouless energy (in
units of the mean level spacing) is the dimensionless conductance,
$g=E_{\rm c}/\Delta$, a parameter which is sufficient to describe
important qualitative features of disordered mesoscopic systems. For
$g \gg 1$ the system is a good conductor, and the eigenfunctions are
delocalized.  Similarly, $g \ll 1$ characterizes an insulator with
localized eigenfunctions.  A transition between both regimes is
observed for $g\approx1$. Moreover, spectral fluctuations for level
spacings smaller than $E_{\rm c}/\Delta$ obey Wigner-Dyson statistics
while for larger level spacings corrections to RMT behavior arise. The
actual value of $g$ depends on the degree of disorder and on the
spatial dimension of the system. In theoretical investigations, $g$ is
usually calculated from an ensemble average. In experiments, however,
it must frequently be obtained from a single sample, i.e.~a single
realization of the disorder potential. The ergodicity hypothesis
requires that both values of $g$ are the same. Due to the observation
that the size of sample-to-sample fluctuations is of the same order as
the fluctuations induced by changes in the energy or magnetic
field~\cite{Lee87}, this hypothesis appears to be valid in such
systems. Since any particular sample has a definite conductance $g$,
which can be measured independent of the existence of other samples,
the ergodicity hypothesis in disordered systems is tantamount to the
statement that the Thouless energy can be extracted from a single
member of the ensemble.

An apparent counter-example to the spectral ergodicity hypothesis has
been found in another class of systems. Complex many-body systems such
as atomic nuclei can be modeled in the framework of random-matrix
theory by the so called two-body random ensemble (TBRE)~\cite{Boh71}.
In this model, fermionic particles are coupled by a stochastic
two-body interaction and distributed over a set of single-particle
states. The many-body Hamiltonian is then represented by a matrix in
the basis of the many-body states constructed as Slater determinants
of the single-particle wave functions. Due to conservation of total
angular momentum, $J$, and total isospin, $T$, the matrix has
block--diagonal structure. More importantly, the two--body nature of
interaction ensures that each block with given $J$ and $T$ is sparse.
Numerical results indicate that the TBRE is non-ergodic in the sense
that spectral and ensemble averages do not
coincide~\cite{Fre71,Flo00}.  One finds the results of spectral
averaging in agreement with experiment~\cite{Bro81}.

A second counter-example to the spectral ergodicity hypothesis can be
found in the spectral fluctuations of the lattice QCD Dirac
operator~\cite{Guh99}. Again, recent developments suggest a connection
to disordered systems. Using formal analogies, it has been argued that
there exists an equivalent of the Thouless energy in
QCD~\cite{Jan98,Osb98a,Osb98b}.  This energy scale can be associated
with properties of the QCD vacuum related to the spontaneous breaking
of chiral symmetry. This resembles the situation in solid states
physics where $E_{\rm c}$ is related to the ground state properties of
the solid, i.e.~$g$. These expectations have been confirmed by
numerical lattice gauge simulations~\cite{Guh99,Ber98}. Moreover,
generalization of the analytic treatments from solid state physics to
include chiral symmetry permits a more rigorous derivation of the
analogies between disordered systems and QCD~\cite{Guh00}. The
essential result is that the spectral fluctuations of the QCD Dirac
operator exhibit RMT behavior up to a certain scale, the equivalent of
a Thouless energy.  Beyond this scale, corrections to the universal
statistics arise. The spectral data that is used to extract the
Thouless energy from Dirac spectra is obtained from an ensemble
average over gauge field configurations. This is the most natural
approach in QCD, since the QCD partition function is defined as a path
integral, i.e.\ an ensemble average. If spectral averaging is
performed nevertheless, an unexpected result is obtained.  The energy
scale seen in the ensemble average of the spectral statistics
disappears, and RMT fluctuations are observed to almost arbitrarily
large scales~\cite{Guh99}. This observed violation of spectral
ergodicity may not be disturbing given the intrinsic nature of
ensemble averaging in QCD. It is currently impossible to say whether
this represents a fundamental difference between QCD and disordered
systems or whether it may be possible to find similar effects in
disordered systems.

We have mentioned three examples of physical systems in
which sparse matrix models apply. In disordered mesoscopic systems the
ergodicity hypothesis is an essential and apparently valid assumption.
The two counter-examples indicate that spectral ergodicity is
not necessarily respected. In each of these cases, however, there are
clear physical arguments indicating which averaging procedure is
appropriate. For the TBRE, the spectral average is natural since one
deals with spectra obtained from single nuclei.  In lattice QCD, on
the other hand, the ensemble average is intrinsic to the system from
its very construction. Nevertheless, it seems important to determine
which spectral properties of these systems are responsible for the
difference between spectral and ensemble averaging and to understand
whether similar effects, disproving the ergodicity hypothesis, can
also be found in disordered systems. One can also imagine new classes
of random matrix models which violate spectral ergodicity but for
which physical arguments in favor of a specific averaging procedure
are not readily available. A better understanding of the mechanisms
responsible for the breakdown of ergodicity could prove to be
essential in such systems. The three examples considered here are, at
first sight, extremely different in character. As noted, their most
obvious common feature is the sparsity of the matrix representations
of their quantum mechanical Hamiltonian operators. Therefore, a
detailed investigation of the influence of sparsity on matrix models
and on the possible breaking of ergodicity in them seems to be
desirable.

Investigations of the TBRE have suggested that the differences between
spectral and ensemble averages are related to fluctuations of the
average spectral density over the ensemble~\cite{Flo00,Fre73}. In this
light, the problem is how to determine the average spectral density.
There is little ambiguity when constructing the ensemble average. By
averaging over a sufficiently large number of members of the ensemble,
it is easy to determine the spectral density in an energy interval of
any given size. The only delicacy in this procedure lies
in the size of the energy intervals chosen. We are interested in
constructing an average local spectral density in the thermodynamic
limit of large matrices. In some cases, such as the Gaussian
ensembles, this double limit of vanishing interval and infinite
particle number is unproblematic. This is not necessarily the case.
Concerning spectral averaging, it is necessary to adopt some kind
of local smoothing. Unfortunately, there is no rigorous and
well defined procedure to accomplish this task, which is usually
regarded as an annoying and cumbersome technical detail. A primary
aim of this paper is, however, to demonstrate that it is precisely the
{\em definition\/} of smoothing which provides the key to understanding
the apparent lack of ergodicity in some systems.

In considering the effect of fluctuations on averaging procedures, it
is essential to recognize that the fluctuations in individual
eigenvalues of a random matrix are not statistically independent.
Indeed, it can be shown that the statistically independent
fluctuations involve the collective motion of literally all
eigenvalues in the spectrum. The statistically independent normal
modes of the spectrum provide a suitable tool for describing this
collective motion~\cite{And98}. As we will show, the nature of these
normal modes is extremely simple, and they can be regarded as plane
waves with a well defined wave length. These normal modes can be
crudely divided into two classes. Short wave fluctuations should
provide information regarding universal spectral properties, and an
appropriate averaging procedure should retain their effects. On the
other hand, long wave length modes describe model-dependent,
non-universal physics and should be eliminated by the averaging procedure.
The art of averaging thus lies in establishing a physically sensible
division between short and long wave lengths. (In this sense, the
challenge of constructing an appropriate spectral average is quite
similar to the task of defining an appropriate energy interval when
performing ensemble averages.) Once this division has been made, the
question of the validity of the ergodic hypothesis can be answered
readily by considering the mean square amplitude of the normal modes
as a function of wave length. In some cases, these mean square
amplitudes have a simple functional dependence on the wavelength,
which applies with equal validity to fluctuations of all wave lengths.
This is the case for the familiar Gaussian ensembles of non-sparse
random matrices, where mean square amplitudes are linearly
proportional to the wave length. In such cases, the details of the
spectral averaging process are irrelevant, and random matrix
ergodicity is respected. In other cases, including the case of sparse
random matrices, the situation is more delicate. There we will
encounter a qualitative difference which, not surprisingly, suggests a
clear distinction between the ``softness'' of long wave length
fluctuations and the relative ``rigidity'' of short wave length modes.
Spectral averaging as commonly employed tends to be more efficient
than ensemble averaging in eliminating these soft long wave length
fluctuations. The result of this argument will provide a simple and
natural explanation of the differences between ensemble and spectral
averaging found in these systems. Further, it will enable us to
understand why the results of spectral averaging show greater
consistency with the familiar results of the non-sparse Gaussian
ensembles. We shall finally see that the apparent breakdown of random
matrix ergodicity is, in fact, a false puzzle. If we ask the same
physical question, the two averaging methods will provide us with the
same answer. If, as can be the case for sparse random matrices,
ensemble and spectral averaging probe different and conflicting
aspects of the system, we should not be startled to obtain different
results.

The paper is organized as follows. We introduce our matrix model in
Sec.~\ref{sec2}.  The relevant parameters of the model are discussed
in Sec.~\ref{sec3}, which is followed by a brief discussion of the
spectral density as a function of these parameters in Sec.~\ref{sec4}.
In Sec.~\ref{sec5}, we present a detailed analysis of the spectral
correlations obtained from the ensemble average. This includes short
range as well as long range correlations. A critical scale is found in
the spectral statistics, which is interpreted in terms of a Thouless
energy. The dependence of this scale on the parameters of the matrix
model is discussed. The findings from the ensemble average are
contrasted with those obtained from spectral averaging in
Sec.~\ref{sec6}. The spectral observables obtained from the two
averaging procedures do not agree. Moreover, the results from spectral
averaging depend on how the local smoothing of the spectral density is
performed. This difference is qualitatively explained by the
collective motion of the eigenvalues. Normal modes are thus introduced
in Sec.~\ref{sec7} to provide a suitable mathematical description of
this collective motion. They will provide us with a natural
explanation of the differences between the results of Sects.~\ref{sec5}
and \ref{sec6} and help us to understand that there is no genuine
violation of random matrix ergodicity. We will offer a summary and
conclusions in Sec.~\ref{sec8}.

\section{Sparse random matrix model}
\label{sec2}

We are interested in some generic features of systems such as
disordered systems, TBRE, and lattice QCD, which can all be described
with sparse random matrices.  Despite the physical differences in
these systems, their spectral properties show similar features.  We
will therefore concentrate on the influence of sparsity on the
spectral properties.  We shall not attempt to model the fine structure
of the TBRE or the multi-band structure of a $d$-dimensional
disordered system.  We will not include the combinatorial correlations
of the TBRE.  We will certainly not attempt to model the complicated
structure of non-abelian lattice gauge theories.

Our matrix model consists of an ensemble of sparse real symmetric
matrices.  We introduce the sparsity $\alpha$, which is the fraction
of the $N(N-1)/2$ independent off-diagonal matrix elements, chosen to
be non-vanishing.  All diagonal elements are kept non-zero.  The
non-vanishing matrix elements $H_{ij}$ are chosen independently and at
random from a Gaussian distribution with mean $0$ and a variance which
is for the diagonal elements $\sqrt{2}\sigma$ and for the off-diagonal
$\sigma$,
\begin{equation}
P(H_{ij})= \frac{1}{\sqrt{2 \pi\sigma_{ij}^2}}
\exp\left(-\frac{H_{ij}^2}{2\sigma_{ij}^2}\right) \ ,
\label{eq22}
\end{equation}
with $\sigma_{ij}^2=1+\delta_{ij}$.  This ensures that GOE behavior is
recovered in the limiting case $\alpha=1$.

Since we are not concerned with the effects of any particular
configuration of the non-vanishing off-diagonal elements, we choose
their positions at random and independently for each matrix of the
ensemble.  We have, however, verified that our results do not change
if one instead maintains randomly chosen but fixed positions for the
non-vanishing elements.

\section{Effective dimension}
\label{sec3}

Our matrix model depends only on two parameters, the matrix size $N$
and the sparsity $\alpha$.  While analytic treatment is difficult for
arbitrary $\alpha$, it is possible to find a qualitative description
of the dependence of spectral correlations on the parameters of the
model.  In particular, comparison with the properties of disordered
systems suggests that there is only one relevant parameter, the
effective dimension.

Consider a particle moving in a $d$-dimensional disordered
medium~\cite{McK81}.  Suppose the strength of the disorder is chosen
so that the motion of the particle is diffusive.  It follows from the
diffusion equation appropriate for the system that the mean square
distance, $\langle x^2\rangle$, traversed by the particle in time $t$ is 
proportional to $t$
\begin{equation}
\langle x^2\rangle=2d{\cal D}t\;,
\label{eq31}
\end{equation}
where ${\cal D}$ is the diffusion constant.  If the system is
restricted to a finite volume $V$, the particle will have explored the
entire system for times larger than the diffusion time. For such times 
the probability of finding the particle at any position $x$ is everywhere
equal. The diffusion time is given by
\begin{equation}
t_{\rm d}=\frac{V^{2/d}}{2d{\cal D}}\;,
\label{eq32}
\end{equation}
which follows from Eq.~(\ref{eq31}) if the mean square distance equals
the linear size of the system, $\sqrt{\langle x^2\rangle}=V^{1/d}$.
The Thouless energy $E_{\rm c}$ is simply the inverse of the diffusion
time $E_{\rm c}\propto1/t_{\rm d}$.  This classical quantity can be
observed in the correlations of the eigenvalues of the quantum
mechanical spectrum.  Spectral correlations are commonly examined on
the scale of the mean level spacing $\Delta$, which scales like
$\Delta\propto1/V$.  Combining the above arguments yields a
dimensionless critical energy, $L_{\rm c}$, given by
\begin{equation}
L_{\rm c}=E_{\rm c}/\Delta\propto d{\cal D}V^{1-2/d}\;.
\label{eq36}
\end{equation}
If the levels have a separation less than $L_{\rm c}$, their correlations
follow the predictions of random matrix theory.  If their separation
is greater than $L_{\rm c}$, corrections to these universal fluctuations
arise.

For $d < 2$, states are localized for any strength of the
disorder~\cite{And58,Mot61}.  The arguments above cannot be applied
since the motion of the particle is {\it not} described by a diffusion
equation, which is an essential assumption in arriving at
Eq.~(\ref{eq31}).  For $d > 2$, states will be delocalized for
sufficiently small disorder. In that case Eq.~(\ref{eq36}) is valid,
and $L_{\rm c}$ can be determined from the spectral statistics.

In order to establish a link to our matrix model, we think of a
suitable discretized lattice version of the corresponding Hamiltonian,
$H$.  In the diffusive regime, the above arguments are also valid for
a lattice realization of the Hamiltonian.  One now expects to find a
critical energy, $L_{\rm c}$, which scales with the volume (now given
by the matrix dimension $N$) in a manner similar to Eq.~(\ref{eq36}).
Thus, we expect that $L_{\rm c}(\alpha,N) \propto {\cal
  C}(\alpha,N)N^{1-2/d(\alpha,N)}$ for some ${\cal C}(\alpha,N)$ and
$d(\alpha,N)$.

For a given $N$ and $d$, it is easy to estimate $\alpha$.  There are
$N(N-1)/2$ independent off-diagonal elements. For each lattice there
are $d$ couplings to neighboring sites in one direction, one for each
dimension. Thus, there are $dN$ non-vanishing matrix elements out of
the total $N(N-1)/2$. The sparsity is then
\begin{equation}
\alpha(N,d)=\frac{dN}{N(N-1)/2}=\frac{2d}{N-1}\;.
\label{eq38}
\end{equation}
Solving for $d$, one obtains an effective dimension $d(\alpha,N)$ for 
any $\alpha$ and $N$.

However, the argument leading to Eq.~(\ref{eq38}) is qualitative, and
several reservations should be mentioned.  Firstly, the coupling of
sites need not be restricted to nearest neighbors, and this can lead
to different factors dependent on the number of couplings. Secondly,
the above arguments are strictly valid only for cubic lattices; other
lattice geometries could lead to additional factors. It remains
nevertheless that the number of couplings grows like $dN$ and the
number of matrix elements grows like $N^2$ for large $N$.  Thus,
$\alpha$ should be proportional to $d/N$.  We therefore introduce an
effective dimension defined as
\begin{equation}
d_{\rm eff}=\alpha N
\label{eq311}
\end{equation}
without additional constants.  By construction, $0\leq\alpha\leq1$ and
thus $0\leq d_{\rm eff}\leq N$.  The effective dimension can also be
understood as twice the average number of independent off-diagonal
elements per row or per column.  For sufficiently large dimensions,
spectral statistics should possess a certain scale $L_{\rm c}$ below
which they follow RMT predictions and above which stronger non-RMT
fluctuations appear.  Thus, we expect that $L_{\rm c}$ should have the
form
\begin{equation}
L_{\rm c}(N,d_{\rm eff})=\left\{
\begin{array}{l@{\quad}l}
c(d_{\rm eff}) N^{\eta(d_{\rm eff})} & d_{\rm eff}>d_{\rm c} \\\\
0 & d_{\rm eff}<d_{\rm c}\;.
\end{array}\right.
\label{eq312}
\end{equation}
For $d_{\rm eff} < d_{\rm c}$, the states should be localized.  In
this region, $L_{\rm c} \ll 1$, and the spectral statistics should be
those of a Poisson distribution.  This is obviously true in the limit
$d_{\rm eff} \to 0$, where the coupling between diagonal elements
vanishes.  The eigenvalues of $H$ are then the randomly distributed
diagonal elements, which obey the Poisson spectral statistics of an
uncorrelated sequence of levels.  In the opposite limit, $d_{\rm eff}
> d_{\rm c}$, states are expected to be delocalized.  Spectral
statistics will have the Wigner-Dyson form up to $L_{\rm c}$.  In the
extreme limit $d_{\rm eff} = N$, $L_{\rm c}$ is equal to $N$ since this
limit corresponds to the original GOE ensemble.  At a certain value
there has to be necessarily a transition between the both limits.  The
exponent should obey
\begin{equation}
\eta(d_{\rm eff})=a-b/d_{\rm eff},
\label{eq313}
\end{equation}
where the constants $a$ and $b$ are independent of $d_{\rm eff}$,
$\alpha$ and $N$.  Since experience does not indicate that spectral
properties are violently sensitive to the sparsity, we expect that
these parameters will be related so that $\eta(d_{\rm c}) \approx 0$.  Thus,
we expect that
\begin{equation}
d_{\rm c} \approx b/a\;.
\label{eq314}
\end{equation}
The constant $c(d_{\rm eff})$ in Eq.~(\ref{eq312}) should depend only
on the effective dimensionality of the system in the case of fixed
disorder strength.  This can be understood from the definition of the
diffusion constant as the second moment of the probability
density~\cite{McK81} in the infinite volume limit, which depends on
the dimension and the degree of disorder.  The latter is kept fixed in
our model.

\section{Spectral density}
\label{sec4}

For arbitrary sparsity, an analytic description of the spectral
density of the matrix model is difficult~\cite{Mir91,Kho97}.  In the
limiting cases $d_{\rm eff}=0$ and $d_{\rm eff}=N$, elementary
analytic expressions are readily available.

\begin{figure}[ht] 
\centerline{\psfig{figure=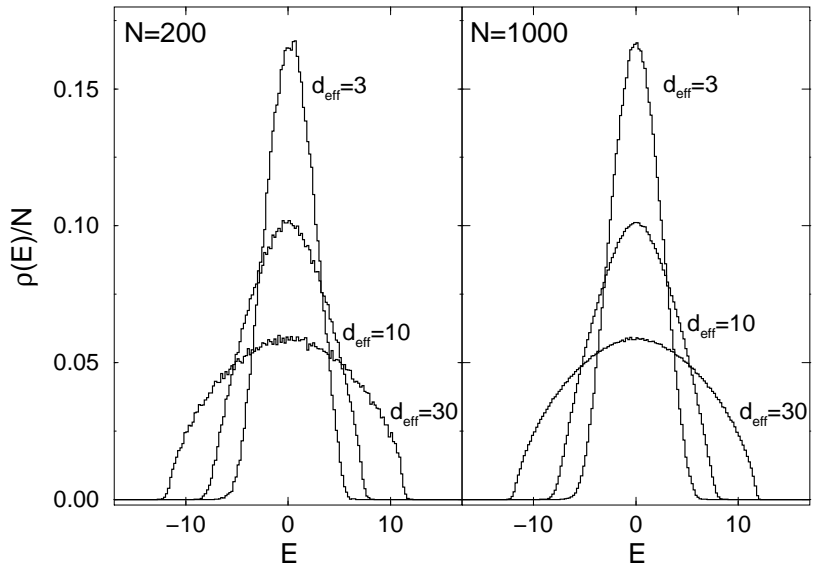,width=0.6\columnwidth,angle=0}}
\caption{The ensemble averaged spectral density, $\rho(E)/N$, for some
$d_{\rm eff}$ and $N$.
\label{fig42}}
\end{figure}

For $d_{\rm eff}=0$, the Hamiltonian matrix reduces to a diagonal
matrix with independent, Gaussian distributed eigenvalues. From
Eq.~(\ref{eq22}) one immediately obtains 
\begin{equation}
\lim_{d_{\rm eff}\to0}\rho(E)
= \frac{N}{2\sqrt{\pi}}\exp\left(-\frac{E^2}{4}\right)\,.
\label{eq41}
\end{equation}
Empirical corrections to the Gaussian shape for small but non-zero
$d_{\rm eff}$ were calculated in Ref.~\cite{Kap00}.  On the other
hand, the matrix model for $d_{\rm eff}=N$ coincides with the standard
Gaussian orthogonal ensemble. The spectral density is then known to
have a semi-circular shape~\cite{Guh98,Kho97},
\begin{equation}
\lim_{d_{\rm eff}\to N}\rho(E) = \frac{1}{2\pi} \sqrt{4 N - E^2}\,.
\label{eq42}
\end{equation}
The spectral density interpolates smoothly between these limits as 
$d_{\rm eff}$ is varied.  The spectral density is conventionally normalized 
to the total number of eigenvalues, $\int_{-\infty}^{\infty} dE\rho(E)=N$.
Fig.~\ref{fig42} shows $\rho(E)/N$ for fixed $d_{\rm eff}$ and varying
$N$.  The ratio $\rho(E)/N$ depends solely on $d_{\rm eff}$. The
overall shape evolves smoothly from an approximate semi-circle to an
approximate Gaussian.  This is consistent with the findings from
many-particle spectra~\cite{Mon75}.

Knowledge of the spectral density is crucial in the statistical
analysis of spectra because spectral correlations must be investigated
on a scale set by the local mean level spacing, $\Delta$, which is
given by $\Delta(E)=1/\rho(E)$.  The transformation
\begin{equation}
\xi=\int_{-\infty}^E dE^\prime\rho(E^\prime)
\label{eq45}
\end{equation}
leads to a dimensionless energy variable, $\xi$, which is the energy
measured in terms of the mean level spacing. This procedure is
referred to as ``unfolding''.  The purpose of this unfolding is to 
eliminate model-dependent macroscopic variations in the spectral 
density in order to reveal underlying universal spectral correlations.

The spectral density can be determined in two ways which differ in 
principle.  The first is to calculate $\Delta$ from an ensemble average. 
The second is to smooth the actual spectral density by local averaging.  
The latter approach is the conventional way of determining $\Delta$
and has been applied to a variety of systems~\cite{Guh98}.  We will
refer to this method as self-unfolding or spectral unfolding.  Obviously, 
when we have only a limited number of experimental samples at our 
disposal, spectral unfolding is the only approach available.  While 
the spectral densities obtained with these methods are superficially 
similar, this can be misleading.  In TBRE~\cite{Flo00} and lattice 
QCD~\cite{Guh99}, these two spectral densities lead to very different 
results.  Ensemble unfolded spectra show a critical scale beyond which 
spectral fluctuations are no longer described by RMT. By contrast, 
self-unfolded spectra show fluctuations which are in excellent agreement 
with RMT expectations on all energy scales.   

It is important to emphasize that these observed differences are due
to the way in which the spectra are unfolded and not to the way in
which spectral correlations within the ensemble of unfolded spectra
are calculated. This fact is frequently ignored and contradicts the
common view that unfolding is a purely technical procedure with no
physical content.  We will discuss this point at some length.

In the following, we denote ensemble averages by a bar,
$\overline{(\ldots)}$.  Spectral averaged quantities will be denoted
by $\langle\ldots\rangle$.  

\section{Spectral observables from ensemble unfolding}
\label{sec5}

In this Section, we investigate several spectral observables using
ensemble unfolding.  We will concentrate on the dependence of these
observables on the effective dimension, $d_{\rm eff}$.  In
Sec.~\ref{sec51}, we study the transition from Poisson to Wigner
statistics exhibited by our matrix model in terms of the nearest
neighbor spacing distribution.  Long range spectral correlations and
the functional dependence of the Thouless energy on $N$ and $d_{\rm eff}$ 
are investigated in Sec.~\ref{sec53}.

\subsection{Short range correlations}
\label{sec51}

The nearest neighbor spacing distribution, $P(s)$, is the probability
density of finding two adjacent levels at a distance
$s=\xi_{i+1}-\xi_i$.  For a sequence of uncorrelated levels, i.e.\ the
Poisson case, this distribution is simply $P_{\rm P}(s)=\exp(-s)$.  In the 
case of random matrices, the nearest neighbor distribution is 
well approximated by the Wigner surmise, which reads $P_{\rm WD}(s) = 
\pi s\exp(-\pi s^2/4)/2$ for real symmetric matrices.

\begin{figure}[ht] 
\centerline{\psfig{figure=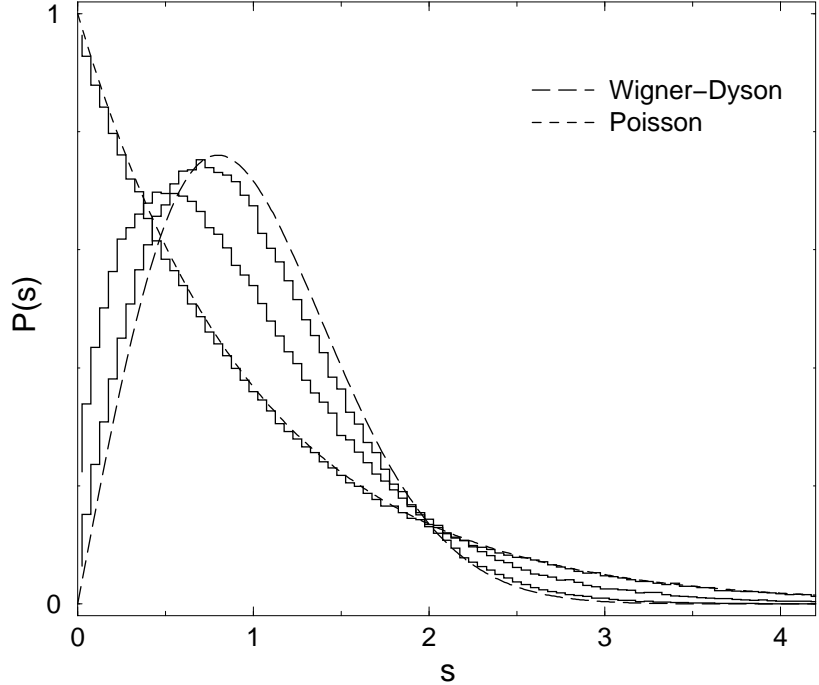,width=0.6\columnwidth,angle=0}}
\caption{The nearest neighbor spacing distribution, $P(s)$, for
  different $d_{\rm eff}$ averaged over an ensemble of 1000 matrices
  of size $N=1000$. The dashed lines correspond to Poisson and
  Wigner-Dyson behavior.  The solid lines correspond to $d_{\rm
    eff}=1,3,5$ and show an evolution from Poisson, $d_{\rm eff}=1$, 
  to almost Wigner-Dyson behavior, $d_{\rm eff}=5$.
\label{fig51a}}
\end{figure}

Fig.~\ref{fig51a} shows the nearest neighbor distribution for fixed
$N$ and varying $d_{\rm eff}$.  When $d_{\rm eff}>4$, the data is
described well by the Wigner surmise.  A transition between Wigner and
Poisson forms is seen in the vicinity of $d_{\rm eff}=2$.  In the
limit of low effective dimension, $P(s)$ is accurately described by
Poisson statistics. This is in agreement with theoretical
considerations~\cite{Mir91}, which suggest that there should be a
transition when the average number of independent non-vanishing
off-diagonal elements is of order one per row.

In order to obtain a more quantitative analysis of the transition of
$P(s)$ observed in Fig.~\ref{fig51a}, we evaluate the integral of the
tail of $P(s)$, $A = \int_{s_0}^\infty P(s) ds$~\cite{Sch93}, which is
free from binning effects, for various values of $d_{\rm eff}$.  Here
$s_0 \approx 2.002$ is the value at which the curves of Poisson and
Wigner form cross. We the Wigner-Dyson and Poisson values as $A_{\rm
WD}$ and $A_{\rm P}$, respectively.  The parameter
\begin{equation}
\gamma(d_{\rm eff}) = \frac{A(d_{\rm eff}) - A_{\rm WD}}{A_{\rm P} -
  A_{\rm WD}} 
\end{equation}
then gives us a quantitative description of the statistics of the spectra
with values $0 \le \gamma \le 1$.  The transition is smooth for all finite 
matrix sizes but becomes sharper as $N$ increases.  A comparison of 
$\gamma(d_{\rm eff})$ for different sizes of the system reveals that all 
curves cross at the same value for the effective dimension
where the size effect changes its sign~\cite{Sch93}.

\begin{figure}[ht]
\centerline{\psfig{figure=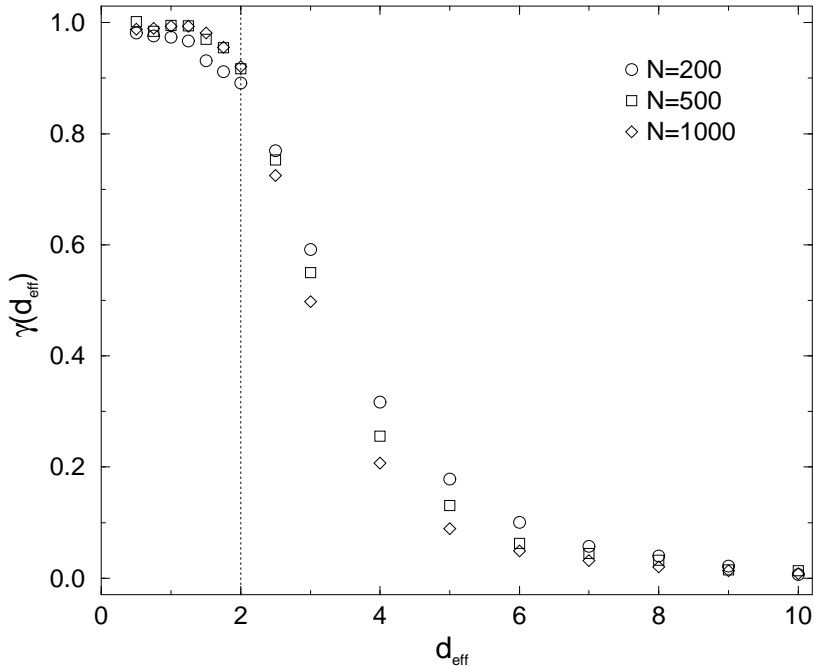,width=0.6\columnwidth,angle=0}}
\caption{$\gamma(d_{\rm eff})$ for three different sizes $N$ of the system:  
$N=200$ (circles), $N=500$ (squares) and $N=1000$ (diamonds).  The
dotted line indicates the critical dimension $d_{\rm eff}=2$
at which the three curves cross. The statistical error is of the order
of the symbol size.}
\label{fig-gamma}
\end{figure}

In Fig.~\ref{fig-gamma}, we plot $\gamma(d_{\rm eff})$ as a function
of the effective dimension for three different values of $N$.  The 
transition from Poisson to Wigner--Dyson spectral fluctuations is
evident. The three curves saturate at the expected values $\gamma = 1$
for small values of $d_{\rm eff}$ and $\gamma = 0$ at large $d_{\rm
eff}$.  As expected, the transition becomes sharper as $N$ increases.
It is also clear from the figure that all curves cross at $d_{\rm eff} 
\approx 2$.  This suggests the existence of a critical dimension for the 
transition between Poisson and Wigner type fluctuations at $d_{\rm eff} 
\approx 2$.

\subsection{Long range correlations}
\label{sec53}

The correlator $P(s)$ and the related parameter $\gamma$ indicate that
there is a qualitative change in the short-range spectral fluctuations
in the vicinity of $d_{\rm eff} = 2$. In this Section, we will extend
this analysis to long-range fluctuations, i.e.\ to scales which are
significantly larger than the mean level spacing.  We do this by
studying the number variance.  All results in this Section have been
obtained from the fluctuations of the ensemble unfolded eigenvalues. A
similar analysis of the self-unfolded eigenvalues and the significant
changes in the properties of long-range fluctuations found there will
be discussed in Sec.~\ref{sec6}.

The number variance measures fluctuations in the total number of
levels found in an energy interval $[L_0-L/2,L_0+L/2]$.  In this
Section $L=\int_{-\infty}^{E}dE^\prime\overline{\rho}(E^\prime)$ is
the ensemble unfolded energy.  The definition of the unfolding process
ensures that the average number of levels in this interval is $L$
independent of $L_0$. The number variance is then given by
\begin{equation}
\overline{\Sigma}^2_{L_0}(L)
=\overline{(n_{L_0}(L)-\overline{n}_{L_0}(L))^2}\;.
\label{eq532}
\end{equation}
The most conservative way to calculate observables such as the number
variance is to perform the ensemble average with fixed energy, $L_0$.
We have, however, verified that our results are independent of $L_0$
provided that we avoid the edges of the spectrum.  This property,
familiar from the Gaussian random matrix ensembles, is called
translational invariance.  Numerical investigation shows that
translational invariance is quantitatively reliable except for some
$10\%$ of the total number of eigenvalues in the vicinity of each of
the edges.  We will restrict our attention to the translationally
invariant bulk of the spectra.  Given translational invariance, we
will omit the subscript $L_0$ in the following.

A sequence of uncorrelated levels, the Poisson case, gives rise to a
strictly linear number variance, $\Sigma^2(L)_{\rm Poisson}=L$. By
contrast, random matrices have much stronger correlations, and the
number variance grows only logarithmically for large $L$. The
asymptotic form for the GOE result for $L \gg 1$ is given by
$\Sigma^2(L)_{\rm GOE}=2/\pi^2 [\ln(2\pi L)+\gamma+1-\pi^2/8]$, where
$\gamma$ is Euler's constant~\cite{Guh98}.

\begin{figure}[ht]
\centerline{\psfig{figure=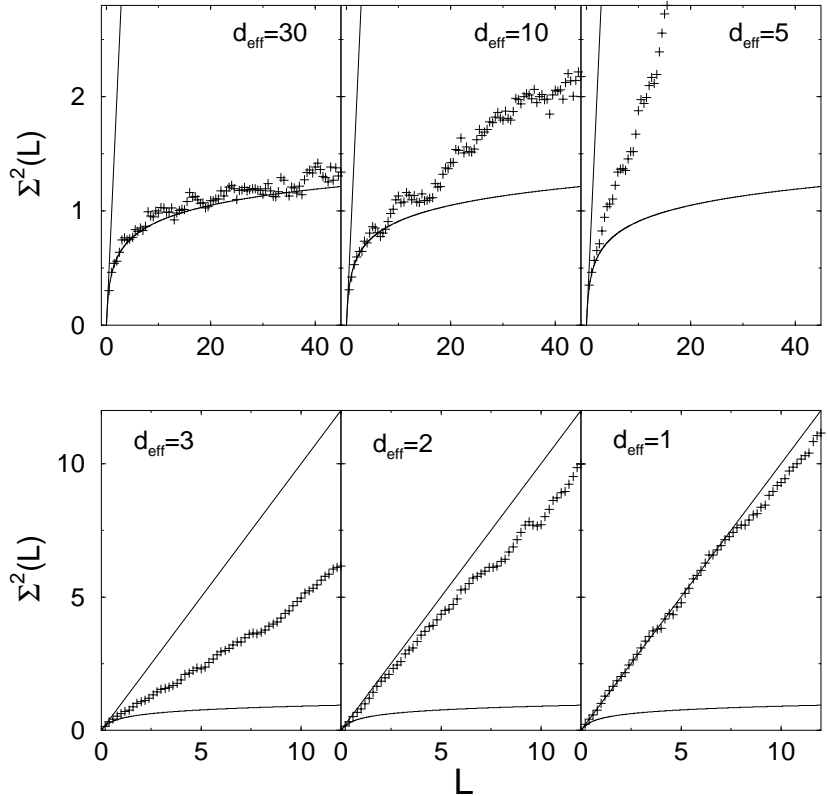,width=0.6\columnwidth,angle=0}}
\caption{The number variance in the central interval $[-L/2,L/2]$ 
for various values of $d_{\rm eff}$.  The matrix size is fixed at
$N=1000$.  The solid lines represent GOE and Poisson results.  Note
that different scales are used in the two rows.
\label{fig53a}}
\end{figure}

Fig.~\ref{fig53a} shows the number variance as a function of $d_{\rm
eff}$.  The matrix dimension is fixed at $N=1000$.  The number
variance is calculated from the central interval $[-L/2,L/2]$ averaged
over the ensemble.  No spectral averaging is performed.  Note that the
same ensemble of matrices was used for the calculation of the number
variance at each value of $L$.  This inevitably results in strongly
correlated statistical uncertainties, which are clearly visible in the
figure.  Above the critical dimension, $d_{\rm c} \approx 2$, the
number variance is described by GOE statistics up to a certain scale
$L_{\rm c}(N,d_{\rm eff})$.  For $L > L_{\rm c}(N,d_{\rm eff})$,
fluctuations become stronger and can no longer be described by the
GOE.  This can be seen in the upper row of Fig.~\ref{fig53a}.  The
critical scale $L_{\rm c}(N,d_{\rm eff})$ decreases as $d_{\rm eff}$
approaches $d_{\rm c}$ from above. The change in the spectral
statistics as one crosses $d_{\rm c}$ is clearly visible in the lower
row of Fig.~\ref{fig53a}.  Well below $d_{\rm c}$, fluctuations show
the purely linear behavior of the Poisson distribution.  This is
similar to the transition seen in the nearest neighbor spacing
distribution of Fig.~\ref{fig51a}.  The critical scale $L_{\rm
c}(N,d_{\rm eff})$ is of order one in the vicinity of $d_{\rm eff}
\approx 2$ and gives rise to the intermediate statistics seen in the
nearest neighbor spacing distribution.

The scale $L_{\rm c}(N,d_{\rm eff})$ is not uniquely defined.  We
elected to use the following definition:
\begin{equation}
\frac{|\Sigma^2(L,N,d_{\rm eff})_{\rm data}-\Sigma^2(L)_{\rm GOE}|}
{\Sigma^2(L)_{\rm GOE}}>\varepsilon \ \ \ {\rm for} \ \ \ 
L>L_{\rm c}^{(\varepsilon)}(N,d_{\rm eff})\;.
\label{eq533}
\end{equation}
The precise value of $L_{\rm c}^{(\varepsilon)}(N,d_{\rm eff})$
depends on the choice of $\varepsilon$.  Moreover, this definition
leads to a non-zero value of $L_{\rm c}^{(\varepsilon)}(N,d_{\rm
eff})$ whenever $\varepsilon>0$ --- even when the data obeys Poisson
statistics. The minimal value of $L_{\rm c}^{(\varepsilon)}(N,0)$ in
that case follows from the solution of Eq.~(\ref{eq533}) with
$\Sigma^2(L,N,d_{\rm eff})_{\rm data}=L$, from which follows that
\begin{equation}
\frac{L_{\rm c}^{(\varepsilon)}(N,0)}{
\Sigma^2(L_{\rm c}^{(\varepsilon)}(N,0))_{\rm GOE}}
=1+\varepsilon\;.
\label{eq534}
\end{equation}
Using the analytic form of the GOE number variance for small
$\varepsilon$, it is easy to show that the solution to
Eq.~(\ref{eq534}) for a pure Poisson distribution is $L_{\rm
c}^{(\varepsilon)} (N , 0) = \varepsilon$ for small $\varepsilon$.
Thus, the observation of $L_{\rm c}^{(\varepsilon)} (N , d_{\rm eff})
\le \varepsilon$ implies that $L_{\rm c} (N , d_{\rm eff}) = 0$.

\begin{figure}
\centerline{\psfig{figure=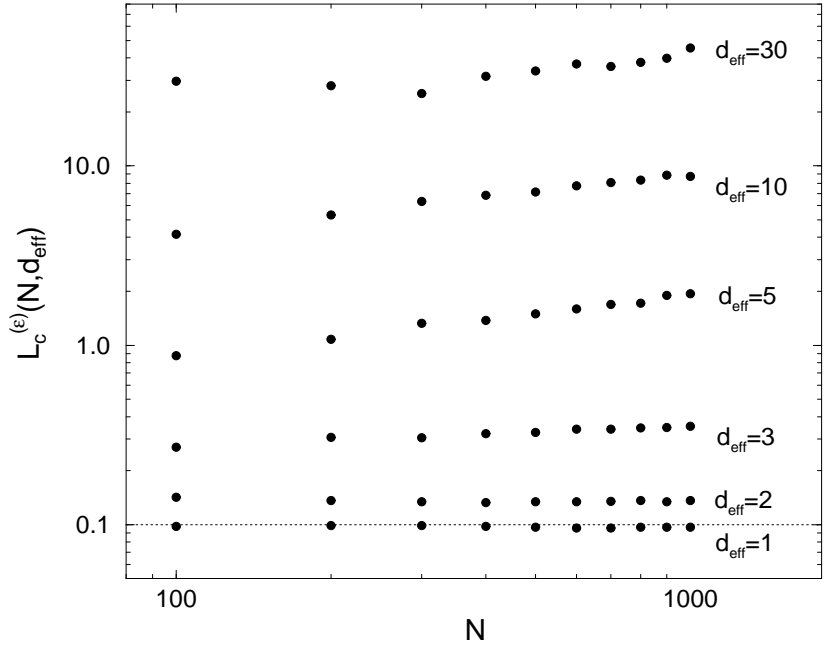,width=0.6\columnwidth,angle=0}}
\caption{The critical scale, $L_{\rm c}^{(\varepsilon)}(N,d_{\rm eff})$, 
  as a function of $N$ for fixed $d_{\rm eff}$ and $\varepsilon=0.1$.
  Changes in $\varepsilon$ affect the overall scale on the $L_{\rm
  c}$-axis, but do not alter the power-law behavior. The dotted line
  corresponds to the Poisson value.
\label{fig53c}}
\end{figure}

The qualitative arguments in Sec.~\ref{sec3} suggest that $L_{\rm c}$
should have a power-law dependence on $N$ for $d_{\rm eff}>d_{\rm c}$.
Both the exponent and the effective diffusion coefficient should
depend only on $d_{\rm eff}$.  Thus, we calculate $L_{\rm
c}^{(\varepsilon)}(N,d_{\rm eff})$ for various $N$ and fixed $d_{\rm
eff}$.  Some examples are shown in Fig.~\ref{fig53c}.  They indicate a
clear power-law dependence on $N$. The value of $L_{\rm
c}^{(\varepsilon)}(N,d_{\rm eff})$ is seen to decrease with decreasing
$d_{\rm eff}$.

\begin{figure}
\centerline{\psfig{figure=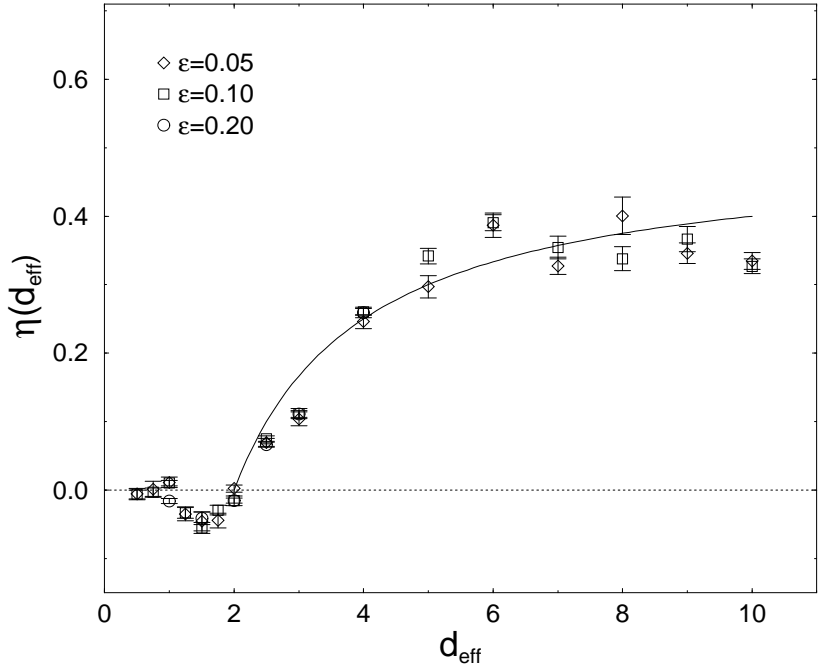,width=0.6\columnwidth,angle=0}}
\caption{The scaling exponent, $\eta(d_{\rm eff})$. The critical dimension
is $d_{\rm c}=2$. The dashed line corresponds to $\eta(d_{\rm eff})=0$ and
the solid line to $\eta(d_{\rm eff})=1/2-1/d_{\rm eff}$.  There is a
transition between the former and the latter at $d_{\rm eff} \approx 2$.
\label{fig53d}}
\end{figure}

\begin{figure}
\centerline{\psfig{figure=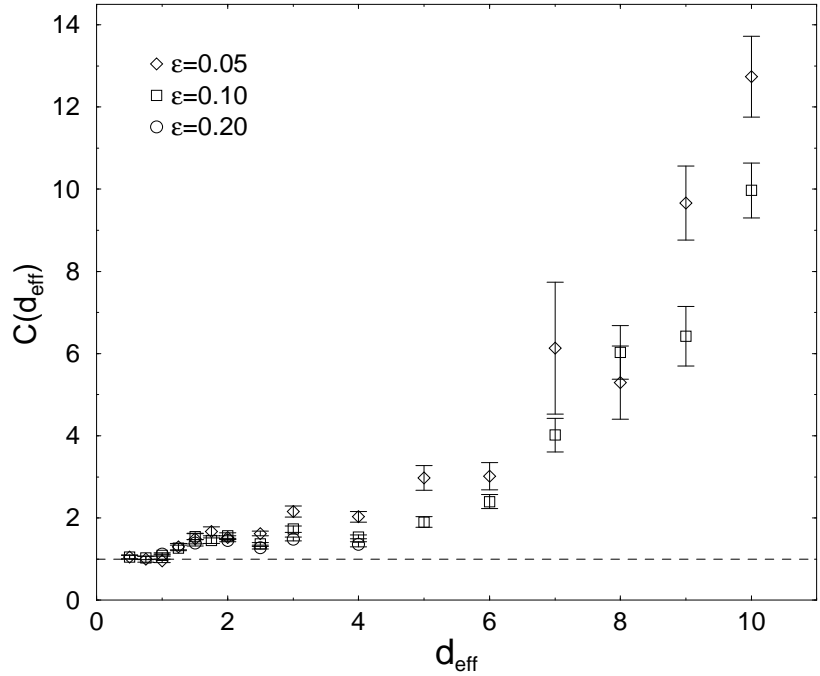,width=0.6\columnwidth,angle=0}}
\caption{Normalized diffusion coefficient 
$C(d_{\rm eff})=c^{(\varepsilon)}(d_{\rm
eff})/c^{(\varepsilon)}(0)$. The dashed line indicates the Poisson
value, $C(0)=1$.
\label{fig53e}}
\end{figure}

The slope in Fig.~\ref{fig53c} is simply the scaling exponent,
$\eta(d_{\rm eff})$, introduced in Eqs.~(\ref{eq312}) and
(\ref{eq313}).  It can be extracted from the data by a linear fit. The
scaling exponent as a function of the effective dimension is shown in
Fig.~\ref{fig53d} as a function of $\varepsilon$.  The error
bars represent the variance of the linear fit.  The exponent
$\eta(d_{\rm eff})$ drops to zero at about $d_{\rm eff}=2$ in
agreement with our expectations.  For $d_{\rm eff} < 2$, the spectral
statistics are of Poisson form, and $L_{\rm c}^{(\varepsilon)}(N,d_{\rm eff}) 
\approx \varepsilon$ independent of $N$ as suggested by Eq.~(\ref{eq534}).  
In contrast to $\eta(d_{\rm eff})$, the diffusion coefficient in
Eq.~(\ref{eq312}), now denoted as $c^{(\varepsilon)}(d_{\rm eff})$,
does depend on $\varepsilon$.  In the Poisson regime, the original
definition of the diffusion coefficient requires $c(0)=0$. However,
given the construction Eq.~(\ref{eq533}), the diffusion coefficient is
non-zero even for $d_{\rm eff}<d_{\rm c}$.  In the limit $d_{\rm eff}=0$, we
have $\eta(0)=0$ and therefore $c^{(\varepsilon)}(0) =
L_{\rm c}^{(\varepsilon)}(N,0)$. We thus normalize the diffusion coefficient
obtained from the linear fit with $c^{(\varepsilon)}(0)$.  The
numerical results are shown in Fig.~\ref{fig53e}.  The normalized
diffusion coefficient is found to be remarkably insensitive to
$\varepsilon$. Thus, we conclude from our matrix model that
\begin{equation}
\frac{L_{\rm c}^{(\varepsilon)}(N,d_{\rm eff})}{L_{\rm c}^{(\varepsilon)}(N,0)}
=C(d_{\rm eff})N^{\eta(d_{\rm eff})}
\qquad d_{\rm eff}>d_{\rm c}\;.
\label{eq535}
\end{equation}
The right hand side of this equation is independent of $\varepsilon$,
which only influences the overall scale.  The scaling exponent
$\eta(d_{\rm eff})$ and the normalized diffusion coefficient $C(d_{\rm
eff}) = c^{(\varepsilon)}(d_{\rm eff})/c^{(\varepsilon)}(0)$ depend
only on the effective dimension.  Below the critical dimension we have
$\eta(d_{\rm eff}<d_{\rm c})=0$. Above the critical dimension, the suggested
functional dependence, Eq.~(\ref{eq313}), is $\eta(d_{\rm eff}>d_{\rm c}) =
a-b/d_{\rm eff}$.  From the analysis of the short range correlators,
we have $d_{\rm c}=2$, which suggests that $a=b/2$.  Performing a one
parameter fit to the data for $d_{\rm eff}>2$, we actually find
$a=0.51\pm0.05$. The function $\eta(d_{\rm eff})=1/2-1/d_{\rm eff}$ is
also shown in Fig.~\ref{fig53d} and is seen to be in reasonable
agreement with the data.


\section{Spectral observables from self-unfolding}
\label{sec6}

We now perform an analysis similar to that of the previous section in
which ensemble unfolding is replaced by self-unfolding
$L=\int_{-\infty}^{E}dE^\prime\langle\rho(E^\prime)\rangle$.  Here,
$\langle\rho(E)\rangle$ is the spectral density obtained from the
local smoothing of each member of the ensemble.  The number variance
is, in analogy to Eq.~(\ref{eq532}), constructed as
\begin{equation}
\langle\Sigma^2(L)\rangle
=\langle\Sigma^2_{L_0}(L)\rangle
=\langle(n_{L_0}-\langle n_{L_0}\rangle)^2\rangle\;.
\label{eq61}
\end{equation}
The average is performed over various intervals $[L_0-L/2,L_0+L/2]$
with fixed interval length, in one spectrum.  The final number variance is 
then obtained as an average over the entire ensemble of matrices.  We note 
that we have employed {\em precisely\/} the same data set in performing 
ensemble and spectral unfolding calculations.  Thus, the resulting 
differences provide a fair measure of differences in the unfolding 
procedure.  

The smoothed spectral density $\langle\rho(E)\rangle$ is not uniquely
defined, and one can think of many methods of equal a priori merit to
obtain it. In our case, we have chosen polynomial unfolding.  We
unfold with a polynomial of fifth order.  The order is kept fixed
throughout the analysis.  Given this unfolding method, an additional
parameter must be fixed, namely the length $L_{\rm fit}$ of the
interval in which we unfold or, equivalently, the fraction $L_{\rm
fit}/N$.

In choosing $L_{\rm fit}/N$, one must be aware of the massive finite
sample size effects that can be encountered.  In a sample of $N$
levels, the number variance vanishes trivially for $L=N$.  The onset
of this effect, however, can already be seen at intervals of length $L
\ll N$.  Numerically studies suggest that interval lengths of $N/10$
or less are required for the GOE in order to avoid finite size
effects, which cause the number variance to {\it decrease\/} beyond a
certain $L$.  By unfolding in a finite interval, $L_{\rm fit}$, the
number of eigenvalues is rather $L_{\rm fit}$ than $N$.  Unwanted
finite size effects can be observed if $L_{\rm fit}$ is chosen too
small.  Such effects are trivial and do not imply a violation of
spectral ergodicity.

\begin{figure}[ht]
\centerline{\psfig{figure=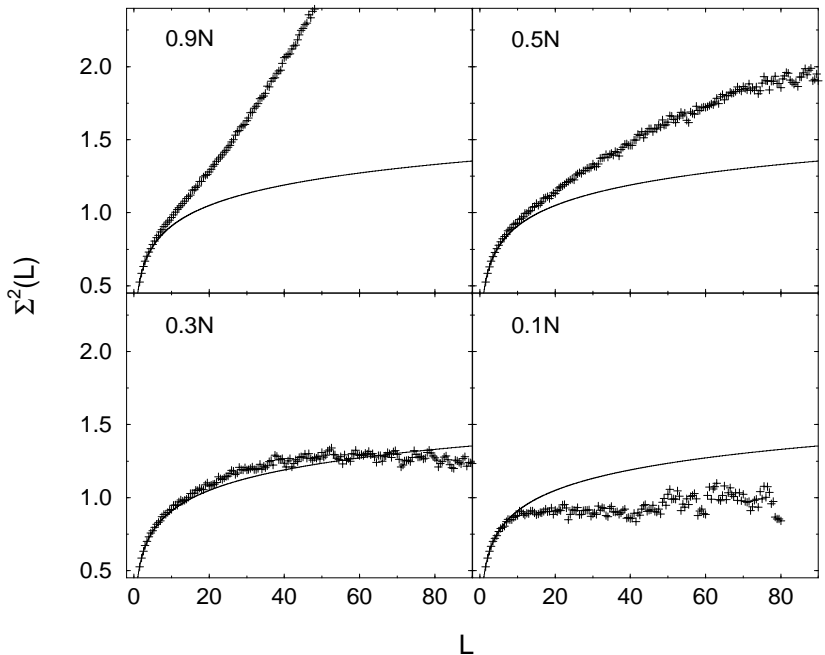,width=0.6\columnwidth,angle=0}}
\caption{The number variance from self-unfolding for 
$d_{\rm eff}=10$ and $N=1000$.  Results are shown for different values of 
$L_{\rm fit}$, given as a fraction of the total number of eigenvalues, $N$.
\label{fig6c}}
\end{figure}

Keeping finite size effects in mind, we proceed to investigate the
effect of $L_{\rm fit}$ on the spectral statistics.  Fig.~\ref{fig6c}
shows the number variance for fixed $N = 1000$ and $d_{\rm eff} = 10$
as a function of $L_{\rm fit}$. For large intervals, $L_{\rm fit}=0.9
N$, the number variance is similar to that obtained from ensemble
unfolding.  Fluctuations are of Wigner-Dyson form for $L\lesssim10$
and then become stronger.  As $L_{\rm fit}$ is reduced from $0.9N$ to
roughly $0.3N$, the onset of the discrepancy between the empirical
results and the Wigner-Dyson is displaced to larger values of $L$.
This is clear from the plot for $L_{\rm fit}/N=0.3$ shown in
Fig.~\ref{fig6c}.  Note, however, the appearance of a saturation of
the fluctuations below the GOE for large values of $L$.  This
saturation is even more pronounced for $L_{\rm fit}/N=0.1$.  The
fluctuations in the case $L_{\rm fit}/N=0.1$ for $L > 10$ are clearly
distorted by finite size effects.  However, the observation remains
that deviations from the GOE are reduced when the length of the
fitting interval is reduced moderately.  The dramatic differences
between the number variance shown here and in the preceding sections
are due strictly to differences in the unfolding procedures.

\begin{figure}[ht]
\centerline{\psfig{figure=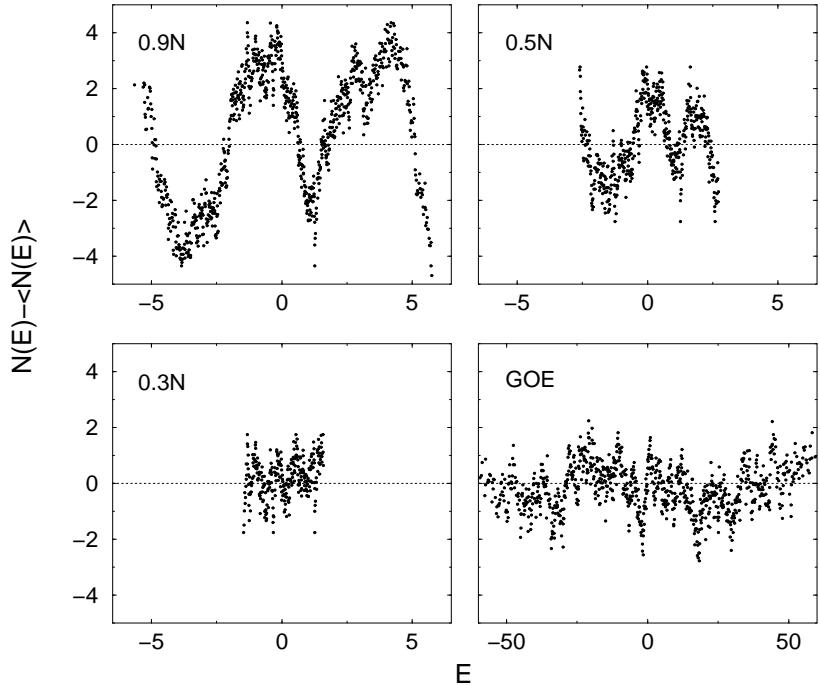,width=0.6\columnwidth,angle=0}}
\caption{The difference between smoothed and actual integrated spectral 
densities for a single matrix with $d_{\rm eff}=10$ and $N=1000$.
Results are shown for $L_{\rm fit}$ equal to $0.9N$, $0.5N$, and
$0.3N$.  The corresponding GOE result is also shown.  Note the different 
scale on the energy axis for the GOE case.
\label{fig6d}}
\end{figure}

A qualitative explanation of the results of Fig.~\ref{fig6c} can be
obtained as follows.  Fig.~\ref{fig6d} shows the difference between
the fitted (i.e.\ smoothed) and integrated spectral densities as a
function of the unfolding interval for a single matrix.  Standard GOE
results are also shown.  While fluctuations are obviously present,
there is also evidence of strong correlations on all energy scales.
Correlations are also clearly present in the GOE results.  The
structure seen on a macroscopic scale is unwanted.  It is in no sense
universal and changes markedly from matrix to matrix.  The goal of
unfolding is thus to remove this non-universal macroscopic structure
while preserving the universal microscopic structure of interest.
This is evidently a delicate task.  If $L_{\rm fit}/N$ is too large,
as for $L_{\rm fit}/N=0.9$, a fixed-order polynomial fit is incapable
of removing all macroscopic structure.  The remaining structure will
lead to a number variance larger than that of the GOE.  If $L_{\rm
fit}/N$ is too small, as for $L_{\rm fit}/N=0.3$, unfolding will begin
to eliminate the genuine correlations which we would like to
investigate.  The resulting number variance will then be smaller than
that of the GOE.  Evidently, there should be some choice of $L_{\rm
fit}/N$ such that the number variance of the GOE is reproduced
essentially exactly.

Although the results of Fig.~\ref{fig6d} are related to the Thouless
energy, they do not provide us with a useful tool for its extraction.
They do provide a qualitative explanation of the differences arising
from the different unfolding approaches.  Ensemble unfolding naturally
includes the effects of fluctuations on all scales and will tend to
maximize the disagreement with RMT.  By contrast, self-unfolding
removes long wave length correlations and gives results for the number
variance which are in better agreement with random matrix theory.
Given its extreme sensitivity to the unfolding procedure, the number
variance cannot be offered as evidence for the violation of spectral
ergodicity.  The safest conclusion is that the choice of unfolding
procedure has important consequences and must be made on the basis of
physical considerations.

\section{Normal mode analysis}
\label{sec7}

We have seen in the previous sections that ensemble and spectral
unfolding do not necessarily lead to the same answers.  Moreover, the
results of the latter depend on the way in which smoothing of the
spectral density is performed.  The differences can be understood from
the collective motion of large numbers of eigenvalues.  The GOE
spectrum is remarkably rigid with regard to correlated fluctuations of
all wave lengths.  While the spectra of sparse random matrices show a
similar stiffness to short wave length fluctuations, they are far more
susceptible to long wave length fluctuations.  The process of
self-unfolding involves the elimination of precisely these long wave
length modes.  Once they have been eliminated, the unfolded spectrum
which remains displays a rigidity much closer to that of the GOE.
Initial investigations along this line were carried out by French, who
associated apparent non-ergodic behavior with the collective motion of
eigenvalues.  He also found a semi-empirical formula which connects
ensemble and spectral averaged number
variance~\cite{Bro81,Fre73}. These findings were recently revisited in
Ref.~\cite{Flo00}.

For our purposes, it is most convenient to consider the set of
``normal modes'' describing the statistically independent fluctuations
of the eigenvalues of a random matrix about their ensemble averaged
locations.  They provide a complete set of functions suitable for the
describing of collective spectral motion \cite{And98}.  We briefly
outline relevant considerations of Ref.\cite{And98}.  The average
positions of the eigenvalues on the unfolded scale is, by
construction,
\begin{equation}
\overline{x_i}=i\;.
\label{eq70}
\end{equation}
The correlation matrix
\begin{equation}
D_{ij}=\overline{x_ix_j}-\overline{x_i}\cdot\overline{x_j}\;,
\label{eq71}
\end{equation}
provides a measure of the fluctuations of the eigenvalues around their
average position.  Evidently, the matrix $D$ is only easy to define from 
an ensemble average.  Since $D$ is a Hermitian matrix, it has $N$ real  
eigenvalues, $\omega_k$, obtained from
\begin{equation}
D_{ij}\psi_j^{(k)}=\omega_k\psi_i^{(k)}\;,
\label{eq72}
\end{equation}
with corresponding eigenfunctions $\psi_i^{(k)}$. The eigenvalues
$\omega_k$ measure the mean square amplitude of the corresponding
fluctuation and are, of course, quite distinct from the eigenvalues,
$x_i$, of the random matrix itself.  By definition, the $\psi_i^{(k)}$
are statistically independent.  The actual positions of the
eigenvalues of any given matrix in the ensemble can always be written
as the sum of their average positions and their fluctuations, $\delta
x_i$, around them.  The fluctuations can be expanded in terms of the
eigenfunctions of $D$, which gives
\begin{eqnarray}
x_i & = & \overline{x_i}+\delta x_i
\nonumber\\
    & = & i + \sum_{k=1}^Nc_k\psi_i^{(k)}\;,
\label{eq73}
\end{eqnarray}
with some coefficients $c_k$.  Given the completeness of the
$\psi_i^{(k)}$, there is a unique correspondence between the $x_i$ and
the $c_k$ for any given matrix.  It is clear that $\overline{c_k}=0$.
Hence, we can express the ensemble number variance Eq.~(\ref{eq532})
as
\begin{eqnarray}
\overline{\Sigma}^2_{L_0}(L) & = &
\overline{(x_{L_0+L/2}-x_{L_0-L/2})^2}-L^2
\nonumber\\ & \approx &
\sum_{k=1}^N\omega_k
\left(\psi_{L_0+L/2}^{(k)}-\psi_{L_0-L/2}^{(k)}\right)^2\;,
\label{eq74}
\end{eqnarray}
where we have made use of the fact that
\begin{equation}
\overline{c_k c_{k^\prime}}=\omega_k\delta_{kk^\prime}\;.
\label{eq75}
\end{equation}
We have made a large $N$ approximation in obtaining the second form in
Eq.~(\ref{eq74}) and have assumed that the normal modes,
$\psi_i^{(k)}$ are smooth functions of $i/N$.  These approximations
lead to acceptable errors of size $1/N$ in
$\overline{\Sigma}^2_{L_0}(L)$.

For the Poisson case with a uniform distribution in the interval
$[0:N+1]$, Eq.~(\ref{eq72}) can be solved exactly using the methods
introduced in Ref.~\cite{And98}
\footnote{This differs slightly from Ref.~\cite{And98}, where 
Eq.~(\ref{eq72}) was solved with the Poisson joint probability density
of the eigenvalues as the starting point.}.  The eigenvalues are
\begin{equation}
\omega_k=\frac{N+1}{4(N+2)
\sin^2\left(\pi k/2(N+1)\right)}
\label{eq78}
\end{equation}
and the corresponding normal modes are
\begin{equation}
\psi_i^{(k)}=\sqrt{\frac{2}{N+1}}
\sin\left(\frac{\pi k i}{N+1}\right)\;.
\label{eq79}
\end{equation}
 
For the GOE, it is convenient to adopt a slightly different approach.  
Instead of constructing the matrix $D$ of Eq.~(\ref{eq72}), it is 
easier to start from the matrix 
\begin{equation}
C_{ij}=\frac{\partial^2}{\partial x_i\partial x_j} 
\log P_N(x_1,\ldots,x_N)\;,
\label{eq711}
\end{equation}
where $P_N(x_1,\ldots,x_N)$ is the joint probability distribution of
the eigenvalues.  The $C$-matrix describes the fluctuations of the
eigenvalues around their average positions.  The corresponding
eigenvalue equation of $C$ can be solved exactly.  The matrices $C$
and $D$ are closely related.  To the extent that the small amplitude
approximation is valid, the eigenfunctions of the matrices $C$ and $D$
are identical; the corresponding eigenvalues are simply reciprocals.
Thus, although $C$ and $D$ are qualitatively similar, they do not
necessarily lead to identical answers.  The eigenvalues of $C$ for the
GOE are thus given by
\begin{equation}
\omega_k=\frac{N}{2\sqrt{2}\pi k}\;,
\label{eq76}
\end{equation}
which is slightly different from the corresponding result in
Ref.\,\cite{And98} due to our use of a different mean level spacing at
$E=0$.  Taking constant factors into account, the normal modes of the
GOE eigenvalues are given in the large $N$ limit by
\begin{equation}
\psi_i^{(k)}=\sqrt{\frac{2}{N}} 
U_{k-1}\left(\frac{\pi (i-N/2)}{N}\right)\;,
\label{eq77}
\end{equation}
where $U_n(x)$ are the Chebyshev polynomials of the second kind. 

With these explicit expressions, the qualitative interpretation of the
normal modes becomes obvious.  The form of the eigenfunctions suggests
that we can roughly associate a wave length, $\sim 1/k$, with each
normal mode.  The normal modes then appear as compressional waves in
the spectra.  In each case, the mode of longest wave length (i.e.\
with $k=1$) involves all eigenvalues moving in the same direction.
The eigenvalues, $\omega_k$, measure the mean square amplitude of the
fluctuations of the various normal modes.  It is clear from
Eq.~(\ref{eq74}) that the largest contributions to the number variance
will come from modes with the largest mean square amplitude.  In the
case of the GOE, this mean square amplitude is strictly proportional
to $1/k$ and provides striking confirmation of the ``rigidity''
associated with the spectra of the Gaussian ensembles.  It is
precisely this feature of the spectrum of normal modes which gives
rise to the logarithmic asymptotic behavior of the number variance.
By contrast, the long wave length normal modes of the Poisson
distribution indicate $\omega_k\approx N^2/\pi^2 k^2$.
Eq.~(\ref{eq74}) now makes it clear that this softness of long wave
length modes is directly responsible for the linear asymptotic
behavior of the number variance.  In short, information regarding the
fluctuations of eigenvalues about their average positions as described
by the normal modes provides a compact source of information regarding
longer range spectral fluctuations.

\begin{figure}[ht]
\centerline{\psfig{figure=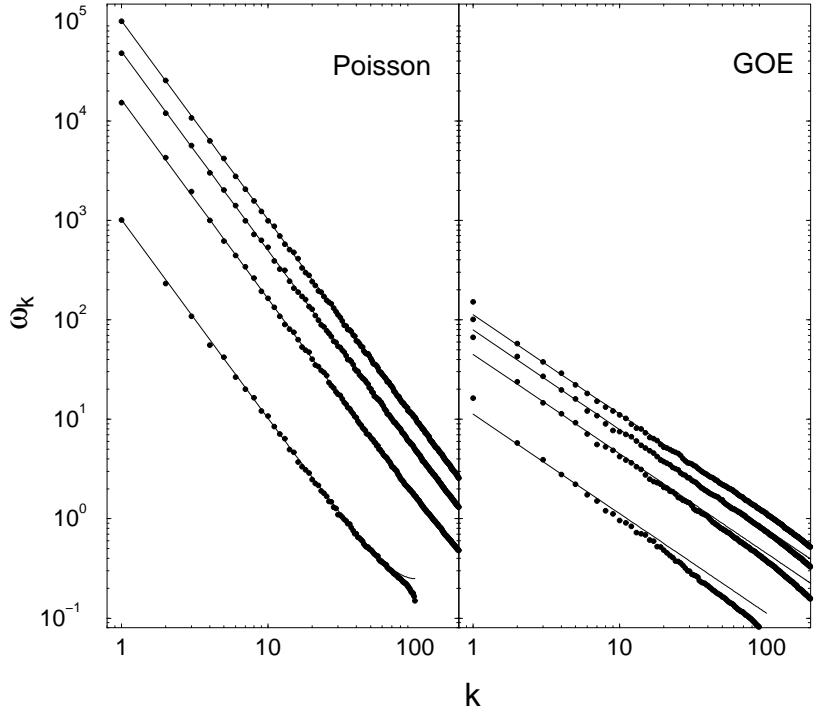,width=0.6\columnwidth,angle=0}}
\caption{The dispersion relations of the soft normal modes from ensembles 
of 1000 matrices for the Poisson, $d_{\rm eff}=0$, and GOE, $d_{\rm
eff}=N$ cases. The solid lines show the
theoretical dispersion relations.  From bottom to top, the data
describe $N=100$, 400, 700 and 1000.
\label{fig7a}}
\end{figure}

We now turn to the numerical solution of Eq.~(\ref{eq72}) with our present 
data and a comparison with the theoretical predictions.  Fig.~\ref{fig7a} 
shows the resulting dispersion relations for both the GOE and Poisson cases.  
In each case, the lowest 100 modes are shown for several matrix sizes.  The 
agreement between theory and numerical data is quite impressive.  In the case 
of the GOE, deviations from the analytical predictions for small matrix 
dimensions are due to the limitations of the Gaussian approximation.

\begin{figure}[ht]
\centerline{\psfig{figure=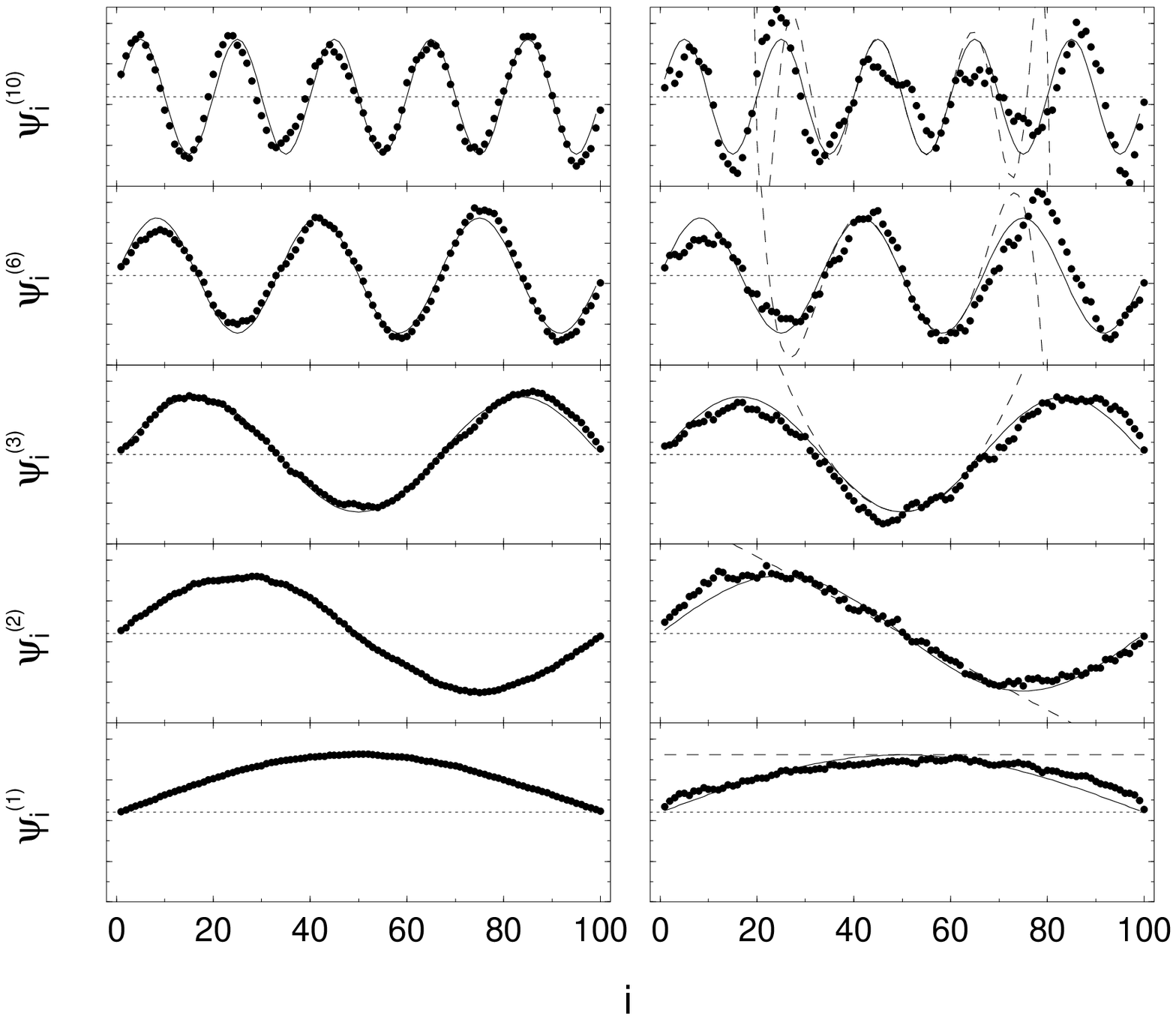,width=0.6\columnwidth,angle=0}}
\caption{Eigenfunctions $\psi_i^{(k)}$ for selected low-lying modes
with $N=100$ for Poisson (left) and GOE (right).  The points
correspond to the data. The solid lines represent Poisson results,
Eq.~(\protect\ref{eq79}).  The dashed lines represent GOE results,
Eq.~(\protect\ref{eq77}). The straight dotted line indicates zero
amplitude.
\label{fig7b}}

The corresponding eigenfunctions of some of the soft modes are shown
in Fig.~\ref{fig7b}.  We have chosen to show the smallest matrix size
of our sets, $N=100$. Qualitative behavior is not altered by a change
of matrix size; only axis scales are different.  Although the
dispersion relations for GOE and Poisson, Fig.~\ref{fig7a}, are
strikingly different and well described by the theoretical
predictions, the normal modes nevertheless follow the Poisson sine
waves, Eq.~(\ref{eq79}).  This is also consistent with theoretical
expectations.  With increasing wave number, $k$, the two theoretical
curves coincide near the center of the spectrum.  This becomes clear
from the explicit expression for the Chebyshev polynomials in terms of
trigonometric functions:
\begin{equation}
U_{k-1}(\cos\theta)=\frac{\sin k\theta}{\sin\theta}\;,
\label{eq710}
\end{equation}
with $\cos\theta=\pi(i-N/2)/N$.  In the center of the spectrum (i.e.\
for $i-N/2\ll N$) and for harder modes (i.e.\ $k\gg1$),
Eq.~(\ref{eq77}) reduces to Eq.~(\ref{eq79}).  The disagreement
between GOE and data is due to the limitations of the Gaussian
approximation.  This approximation is only capable of describing small
amplitude fluctuations.  As a result, the large amplitude motion
associated with the soft, long wave length mode is not described
quantitatively with this approximation.  This limitation will not
influence our analysis significantly.  The determination of the
asymptotic form of the number variance requires that we first take the
limit $N \to \infty$ and then take the limit $L \to \infty$.  If $L_0
= 0$, this requires knowledge of the normal modes in the middle of the
spectrum, where the agreement between data and theoretical
expectations is satisfactory.

Figs.~\ref{fig7c} and \ref{fig7b2} show the dispersion relations and
wave functions of the normal modes for some intermediate cases, $0 <
d_{\rm eff} \ll N$.  The wave functions for the normal modes are again
reasonably well described by plane waves, and it thus makes sense to
interpret the eigenvalue number, $k$, as a wave number.  The resulting
dispersion relations are clearly no longer scale invariant.  The
character of the spectrum changes qualitatively from a long wave
length $1/k^2$ behavior, as seen above in the Poisson distribution, to
the short wave length $1/k$ behavior previously encountered in the
GOE.  The nature of the spectrum changes abruptly at a critical wave
number, $k_{\rm c}$.  Numerical investigations suggest, for example,
that $k_{\rm c}$ scales with $\sqrt{N}$ for $d_{\rm eff} \approx 10$.
Thus, the soft modes with $k < k_{\rm c}$ represent a vanishing
fraction of the normal mode spectrum in the thermodynamic limit.  The
precise value of $k_{\rm c}$ depends sensitively on the value of
$d_{\rm eff}$ considered.

\begin{figure}[ht]
\centerline{\psfig{figure=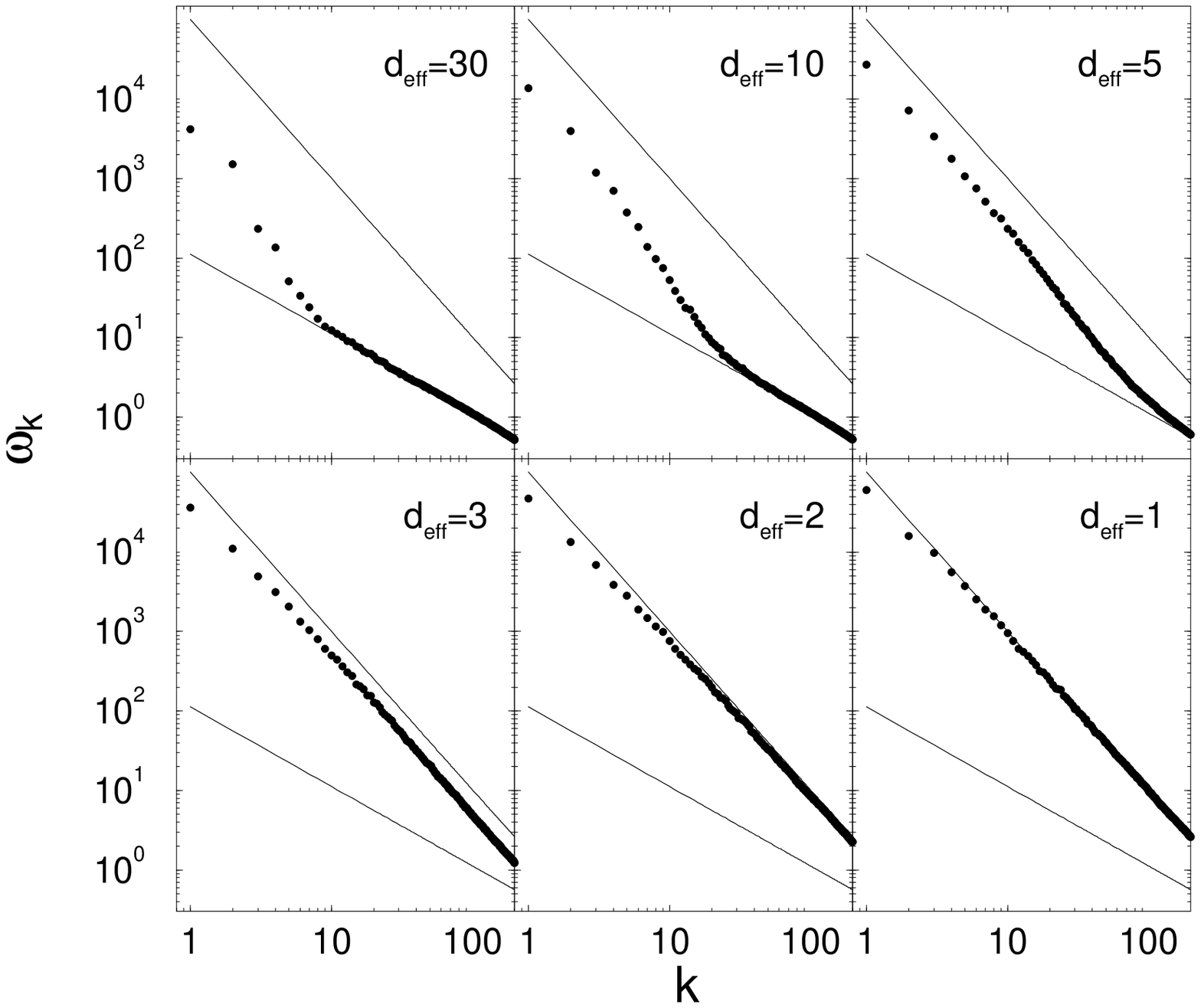,width=0.6\columnwidth,angle=0}}
\caption{The dispersion relation of the soft normal modes for various 
values of $d_{\rm eff}$ as calculated from an ensemble of 1000
matrices of dimension $N=1000$.  The upper solid line corresponds to
the pure Poisson case; the lower solid line corresponds to the pure
GOE.\label{fig7c}}
\end{figure}
\end{figure}

\begin{figure}[ht]
\centerline{\psfig{figure=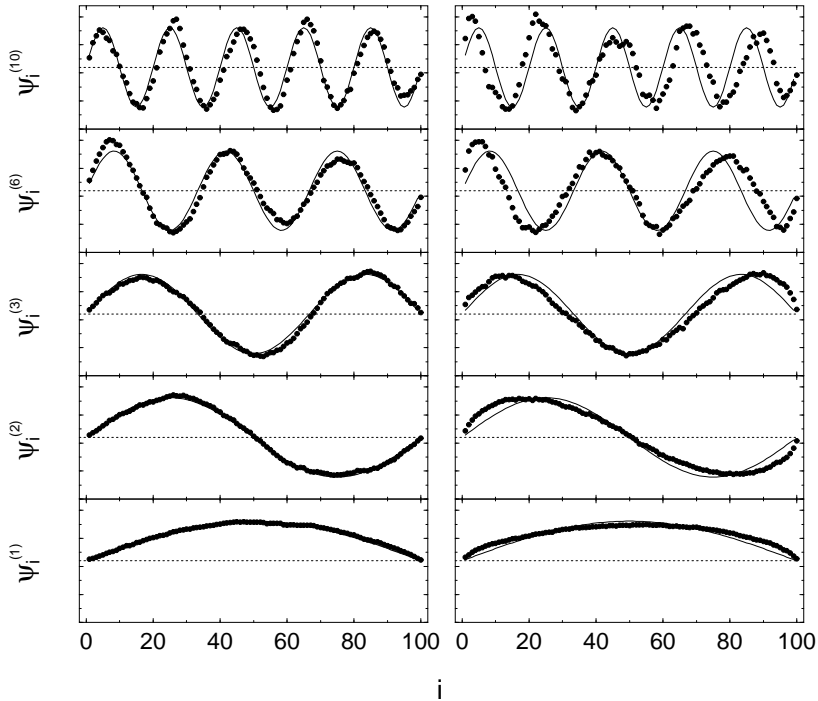,width=0.6\columnwidth,angle=0}}
\caption{Same as Fig.~\protect\ref{fig7b} but for the cases $d_{\rm eff}=2$ 
(left) and $d_{\rm eff}=10$ (right).
\label{fig7b2}}
\end{figure}

These results lead immediately to the important observation that the
eigenfunctions of $D$ for arbitrary $d_{\rm eff}$ are reasonably well
described by the sine waves of Eq.~(\ref{eq79}) independent of whether
the spectral statistics are pure GOE, pure Poisson, or a
scale-dependent mixture of the two.  As noted, the eigenvalues,
$\omega_k$, depend sensitively on the statistics chosen (i.e.\ on
$d_{\rm eff}$).  It is therefore reasonable to approximate the number
variance at $L_0 = 0$, according to Eq.~(\ref{eq74}), as
\begin{equation}
\overline{\Sigma}^2(L) \approx 
\frac{4}{N} \, \sum_{k=1}^N\omega_k \sin^2\left(\frac{\pi L k}{2 N}\right)\;.
\label{eq712}
\end{equation}
In this form, we see that the number variance is largely determined by the 
dispersion relation of the eigenvalues $\omega_k$.  In particular, the large 
$L$ behavior is dominantly set by the softest normal modes.

The normal mode spectra and Eq.~(\ref{eq712}) can now provide us with
a simple explanation of the differences between the number variance
obtained from ensemble and spectral unfolding in the case of matrices
with sparsity $0 < d_{\rm eff} \ll N$.  Let us first consider the case
of ensemble unfolding. First, consider $L$ to be of order $1$.  The
contribution of long wave length modes, for which $\omega_k \sim 1/k^2$, is
suppressed by the factor $k^2$ coming from the sine term.  The
contribution from short wave length modes (i.e.\ $k$ of order $N$),
for which $\omega_k \sim 1/k$, persists and builds the logarithmic
behavior of the number variance familiar from the GOE.  As $L$
increases, the long wave length modes are no longer suppressed.  Their
$1/k^2$ contributions now build up the linear divergence of the number
variance familiar from the Poisson distribution.  The resulting qualitative 
behavior is precisely that shown in Fig.~\ref{fig53a}.

\begin{figure}[ht]
\centerline{\psfig{figure=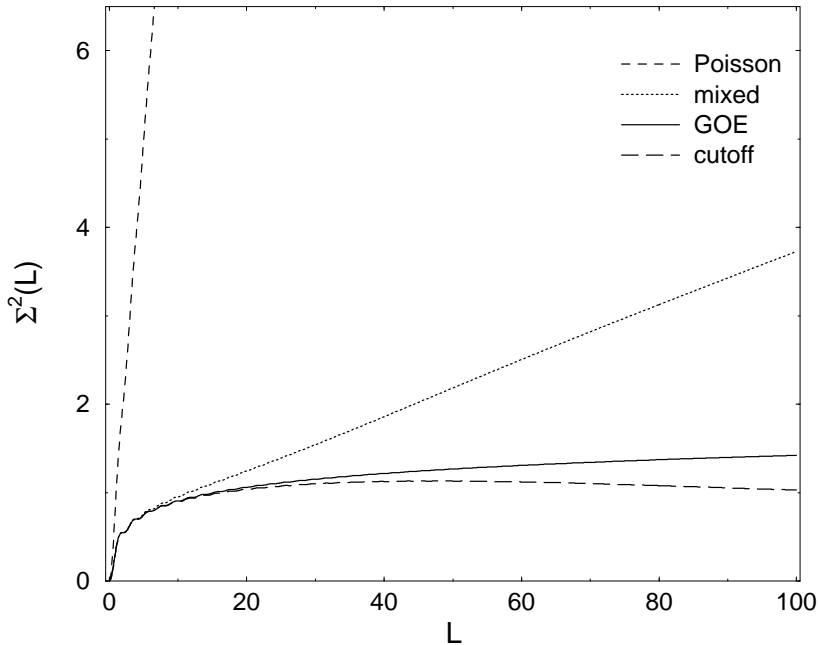,width=0.6\columnwidth,angle=0}}
\caption{
Number variance calculated from the sum in Eq.~(\protect\ref{eq712})
with $N=1000$.  The different dispersion relations $\omega_k$ are
Eq.~(\protect\ref{eq78}) (Poisson), Eq.~(\protect\ref{eq76}) (GOE),
Eq.~(\protect\ref{eq713}) with $k_{\rm c}=\protect\sqrt{N}$ (mixed),
and Eq.~(\protect\ref{eq714}) with $k_0=10$ (cutoff).
\label{fig7d}}
\end{figure}

In Fig.~\ref{fig7d}, we give some examples of the number variance
obtained from evaluating the sum Eq.~(\ref{eq712}). We consider the GOE,
Poisson, and two specific toy dispersion relations motivated by the results 
of our sparse matrix model.  The first of these is intended to mimic the 
results of ensemble unfolding.  We construct a dispersion relation 
which interpolates between GOE and Poisson forms, i.e.
\begin{equation}
\omega_k=\left\{
\begin{array}{lc}
\omega_k^{\rm (Poisson)}/k_{\rm c} & k\leq k_{\rm c} \\
\omega_k^{\rm (GOE)}  & k>k_{\rm c}
\end{array}\right.\;,
\label{eq713}
\end{equation}
where $\omega_k^{\rm (Poisson)}$ and $\omega_k^{\rm (GOE)}$ are given
by Eqs.~(\ref{eq78}) and (\ref{eq76}), respectively.  The choice
$k_{\rm c} = \sqrt{N}$ leads to a nearly continuous dispersion
relation and a surprisingly faithful reproduction of the dispersion
relation shown in Fig.~\ref{fig7c} for the case $d_{\rm eff} = 10$.
The resulting number variance is in striking agreement with the
ensemble average results shown in Fig.~\ref{fig53a}.

The second toy dispersion relation is intended to demonstrate the
striking sensitivity of the number variance to the treatment of the
very softest normal modes through the introduction of a simple cutoff
\begin{equation}
\omega_k=\left\{
\begin{array}{lc}
0  & k\leq k_0 \\
\omega_k^{\rm (GOE)}  & k>k_0
\end{array}\right.\;.
\label{eq714}
\end{equation}
It is evident from Eq.~(\ref{eq712}) that every normal mode makes a 
positive contribution to $\overline{\Sigma}^2(L)$.  The elimination 
of soft modes resulting from the use of Eq.~(\ref{eq714}) will 
necessarily lead to a number variance everywhere smaller than the GOE 
result.  Even the modest choice of $k_0 = 10$ for $N=1000$ shown in 
Fig.~(\ref{fig7d}) is sufficient to cause the number variance to saturate 
and decrease for sufficiently large $L$.

These simple examples lead us towards a better understanding of the
results of self-unfolding.  As we have seen, it is necessary to
eliminate non-universal, long wave length fluctuations in the spectrum
(i.e.\ unfold the spectrum) if we are to be able to make sensible
comparisons of spectral fluctuations in macroscopically distinct
regions of the spectrum.  Let us now consider unfolding from the point
of view of Eq.~(\ref{eq73}).  The eigenvalues from any given
realization of a random matrix uniquely determine the expansion
coefficients, $c_k$, in the complete set of normal modes.
Self-unfolding then corresponds to imposing a smooth cutoff on these
coefficients to eliminate the contribution of normal modes with small
$k$.  In other words, one replaces $c_k$ by $G(k) c_k$ where $G(k) \to
0$ as $k \to 0$.  Obviously, the scale on which $G(k)$ vanishes must
be set by physical arguments.  If desired, of course, one can invert
Eq.~(\ref{eq73}) and obtain the resulting unfolded eigenvalues.
Qualitatively, one can proceed directly to the evaluation of the
number variance through the approximate Eq.~(\ref{eq712}).  Using
Eq.~(\ref{eq75}), it is clear that the primary effect of the unfolding
just described is to replace the terms $\omega_k$ by $\omega_k G^2
(k)$.  This has the effect of greatly reducing the contribution of
long wave length modes to the number variance.  However, these are
precisely the modes which are responsible for the large $L$ form of
the number variance.

There are now several possibilities.  If the normal mode spectrum is
scale invariant, the suppression of long wave length modes implied by
unfolding will have little consequence.  The number variance will have
the same qualitative behavior independent of the unfolding method
adopted.  This is the case for both the Poisson distribution and the GOE.
The situation is quite different for sparse random matrices where the 
normal mode spectrum displays a transition from Poisson to Gaussian 
form on a scale $k_c \sim \sqrt{N}$.  The number variance obtained 
from ensemble unfolding will show deviations from random matrix theory.  
If the scale of $G(k)$ implicit in self-unfolding is sufficiently large 
to eliminate the contributions of soft normal modes, we will obtain the 
logarithmic behavior of the Gaussian ensembles.  While such disagreement 
has been viewed as a breaking of spectral ergodicity, this is not the 
case.  Rather, ensemble and spectral unfolding explore the same normal 
mode spectra at different scales.  If different questions are asked in the 
two cases, we should not be surprised to get different answers.  Although 
it is conventional to view self-unfolding as the more ambiguous procedure, 
it is also possible to restore agreement between the two methods by modifying 
the ensemble unfolding procedure.  Specifically, this could be accomplished 
by increasing the bin size used in Eq.~(\ref{eq45}) so that each bin included 
roughly $\sqrt{N}$ levels.

\begin{figure}[ht]
\centerline{\psfig{figure=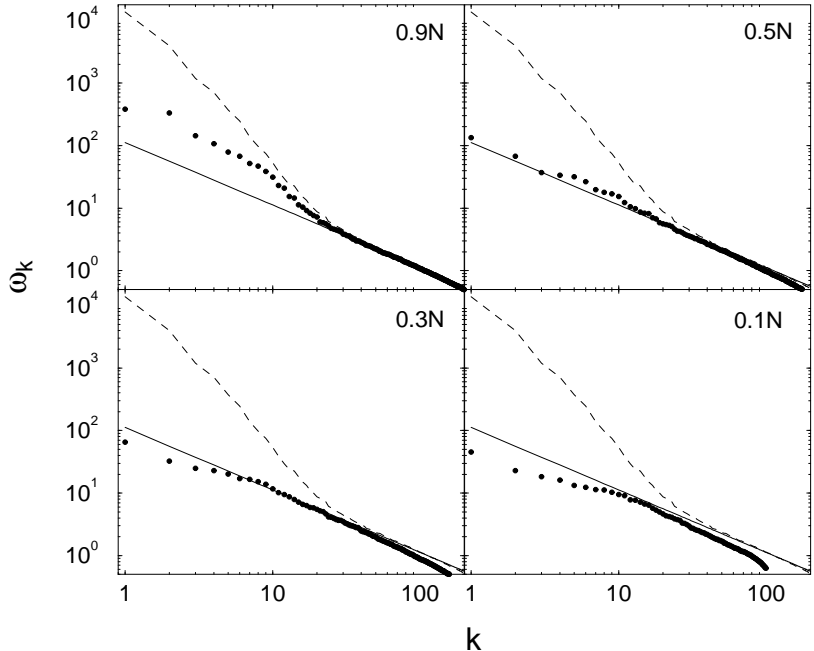,width=0.6\columnwidth,angle=0}}
\caption{Normal mode spectrum after self-unfolding for $N=1000$
and $d_{\rm eff}=10$, cf.~Fig.~\protect\ref{fig6c}.  The solid line
corresponds to GOE and the dashed line to the ensemble unfolded
result.
\label{fig7e}}
\end{figure}

The effect of polynomial self-unfolding on the normal modes can be
seen in Fig.~\ref{fig7e}.  Here, we construct the matrix $D$ according
to Eq.~(\ref{eq71}) using the self-unfolded eigenvalues.  The normal
modes in Fig.~\ref{fig7e} are obtained from the same data used in
constructing Fig.~\ref{fig7c}.  The effects are quite dramatic but
entirely predictable.  The purpose of unfolding is to eliminate or
greatly reduce the mean square amplitude of long wave length normal
modes (i.e.\ reduce $\omega_k$ for small $k$.)  This figure reveals
that the softest modes are lowered by one order of magnitude for an
unfolding interval of length $L_{\rm fit} = 0.9N$.  For $L_{\rm fit}$
of $0.4N$ to $0.5N$, the relation dispersion is remarkably similar to
that of the GOE, which is consistent with our results for $\Sigma^2$.
For even smaller intervals, the dispersion relation falls below the
GOE and gives rise to the saturation effects seen in Fig.~\ref{fig6c}.
It is clear that the effective cutoff, $G(k)$ can be determined
immediately as the ratio of the results of Fig.~\ref{fig7e} to those
of Fig.~\ref{fig7c}.

\section{Summary and conclusions}
\label{sec8}

We have presented a detailed analysis of the spectral properties of
sparse random matrix ensembles with particular emphasis on the
spectral ergodicity hypothesis and the effects of sparsity on spectral
statistics.  In addition, we have presented the first numerical
investigation of a normal mode spectrum in random matrix ensembles.
The question of spectral ergodicity is of direct relevance for a
variety of physical systems.  The existence of apparent counter
examples, such as lattice QCD and many-body systems with two-body
interactions, suggests that this hypothesis is not always fulfilled.
The correct averaging procedure seems clear in these two cases.  In
lattice QCD, the ensemble average is the relevant procedure by
construction.  Spectral averaging seems physically meaningless, and
differences between the approaches have not been regarded as
surprising.  On the other hand, the spectral averaging is evidently
more appropriate for the TBRE, since we are interested in an ``average
nuclear spectrum'' and not a spectrum averaged over many distinct
nuclei.  In each of these cases, the physically motivated averaging
procedure agrees with data.  In the case of disordered systems, the
ergodicity hypothesis is a crucial ingredient necessary for the
comparison of theory and experiment.  Due to the similarities of all
three systems, e.g.\ with respect to spectral statistics, it is
important to understand if disordered systems can show non-ergodic
properties or if, on the other the hand, TBRE and lattice QCD theory
are in reality ergodic.  Our analysis indicates the latter and
suggests that apparent differences between spectral and ensemble
averaged results have been misinterpreted.

Using ensemble unfolding, we presented a detailed analysis of the
parameter dependence of our model. For a certain parameter range,
spectral correlations are described by RMT; outside this range,
fluctuations are stronger. The dependence of this critical energy
scale on the model parameters can be understood from viewing the model
as a multi-dimensional disordered system on a random lattice with
fixed disorder strength.  Results obtained from a self-unfolding of
the spectra do not agree with the results of ensemble unfolding.
There is also a strong dependence on the parameters of the
self-unfolding procedure, i.e.\ the length of the unfolding interval.
Changes of this length can alter the critical energy obtained with
ensemble unfolding.  Indeed, this length can even be chosen to
reinstate virtually perfect agreement with the spectral correlations
of random matrix theory.

The observed differences between the results of ensemble and spectral
unfolding have been shown to be a result of the collective motion of
large numbers of eigenvalues.  Such differences are intrinsic to
systems lacking scale invariance, for which correlations change from
RMT to Poisson on some scale.  Ensemble unfolding includes the effects
of fluctuations on all scales.  Self-unfolding necessarily eliminates
the effects of ``soft'', long wave length fluctuations, whose
contributions to the asymptotic number variance would otherwise be
dominant.  We have shown that the results of self-unfolding are likely
to be closer to those of RMT and are likely to differ from the results
of ensemble unfolding.  Spectral ergodicity is not broken.  Rather,
the two approaches probe different scales of the spectrum.

The normal modes of the random matrix proved to be a convenient tool
for the investigation of spectral correlations.  They describe the
fluctuations of the eigenvalues around their average positions and can
be understood approximately as compressional waves.  Our analysis
shows that the normal modes are well described as plane waves,
independent of the parameters of our model.  Their dispersion
relation, however, is sensitive to the choice of parameters.  Spectral
correlations are therefore largely determined by dispersion relation,
as indicated by the explicit expression for the number variance given
above.  The presence of a clear scale dependence in the dispersion
relation for sparse random matrices was sufficient to provide a
qualitative explanation of the apparent violation of spectral
ergodicity in this problem.  This example serves a striking reminder
that the unfolding procedure, usually regarded as technically
difficult and uninteresting, can have important physical content.  An
appropriate unfolding procedure should reflect the spectral scale
which is relevant for the physical properties in question.  This scale
is not always apparent given the usual ad hoc treatment of unfolding.
The normal modes permit a more systematic formulation of this problem.

Given the similarities between disordered systems and the sparse
random matrices considered here and the matrices appropriate for
disordered systems, it would be of interest to perform a normal mode
analysis appropriate for this case as well.  A scale invariant
dispersion relation could provide strong confirmation of spectral
ergodicity in disordered systems; the presence of a scale could
indicate interesting new lines of experimental investigation.

\section*{Acknowledgments}
We are grateful to acknowledge discussions with T. Seligman and
T. Papenbrock.

\end{document}